\documentclass[a4paper,11pt]{article}
\pdfoutput=1 

\usepackage{jcappub} 

\usepackage[T1]{fontenc} 

\title{\boldmath Possible relationship between  initial conditions for inflation
\\
and past geodesic incompleteness of the inflationary spacetime}

\author{ Alexander  B. Kaganovich}

\affiliation{Physics Department, Ben Gurion University of the Negev \\
Beer Sheva, Israel\\
Sami Shamoon College of Engineering,\\ 
Beer Sheva, Israel}

\emailAdd{alexk@bgu.ac.il}

\abstract{According to the Borde-Guth-Vilenkin (BGV) theorem
an expanding region of spacetime cannot be extended to the past beyond some boundary $\mathcal{B}$.
Therefore, the inflationary universe must have had some kind of beginning. However, the BGW theorem says nothing about the boundary conditions on $\mathcal{B}$, or even about its location.
Here we present a single-scalar field model of the Two-Measure Theory, where the non-Riemannian volume element $\Upsilon d^4x$ is present in the action.
As a result of the model dynamics, an upper bound $\varphi_0$ of admissible  values of the scalar field $\varphi$ appears, which sets the position of $\mathcal{B}$ in the form of a spacelike hypersurface $\Upsilon(x)=0$ with a  boundary condition: $\Upsilon\rightarrow 0^+$ as $\varphi\rightarrow\varphi_0^{\,\,\,-}$.
A detailed study has established that if the initial kinetic energy density  $\rho_{kin}^{(in)}$ prevails over initial gradient energy density $\rho_{grad}^{(in)}$ then there is an interval of initial values $\varphi_{in}^{(min)}\leq  \varphi_{in}
<\varphi_0$, where  $\rho_{kin}^{(in)}$ and $\rho_{grad}^{(in)}$ cannot exceed the potential energy density and hence  the initial conditions necessary for the onset of inflation are satisfied.
It is shown that under almost all possible  left-handed boundary conditions on $\mathcal{B}$, that is where $\Upsilon\rightarrow 0^-$, the metric tensor in the Einstein frame has a jump discontinuity on $\mathcal{B}$, so the Christoffel connection coefficients are not defined on the spacelike hypersurface $\Upsilon=0$. Thus,
if $\varphi_{in}^{(min)}\leq \varphi_{in}<\varphi_0$ and $\rho_{kin}^{(in)}>\rho_{grad}^{(in)}$, then there was an inflationary stage in the history of our Universe and
the congruence  of timelike geodesics cannot be extended to the past beyond the hypersurface $\Upsilon=0$.}

\begin{document}
\maketitle
\flushbottom

\section{Introduction}
\label{sec:intro}

The purpose of this paper is to present a simple model in which, over a wide range of parameters, there is {\em a close relationship between two fundamental problems of inflationary cosmology}. One of them is the problem of the initial conditions necessary for the onset of inflation \cite{singular alpha}.
The second problem is related to the Borde-Guth-Vilenkin (BGV)  theorem \cite{BGV}, which states that inflationary models cannot be extended to the past beyond some boundary $\mathcal{B}$. As will be shown, this relationship allows us to get answers to the questions that are the main content of these two problems.

\begin{itemize}

\item

{\em The problem of initial conditions for inflation.}

  Conditions that the initial kinetic and gradient energy densities of the canonically normalised scalar field  should not exceed the potential energy density 
\begin{equation}
\rho_{kin}^{(in)}= \frac{1}{2}\dot{\varphi}_{in}^2\lesssim V(\varphi_{in}) \quad \text{and} \quad  \rho_{grad}^{(in)}=\frac{1}{2}|(\partial^k\varphi)_{in}(\partial_{k}\varphi)_{in}|\lesssim V(\varphi_{in})
\label{Cond for begin infl}
\end{equation}
are well known as the constraints neeeded for the onset of inflation.  According to the understanding developed in the first models of chaotic inflation \cite{Linde 1983}, \cite{Linde 1985}, when the classical space-time domain first appears after the Planck quantum era, the total energy density is of the order of $M_P^4$, and inflation begins with $V(\varphi_{in})\sim M_P^4$. Then all admissible values $\varphi_{in}$, $\dot{\varphi}_{in}$, $(\partial_{k}\varphi)_{in}$ of the classical scalar field $\varphi$ satisfying (\ref{Cond for begin infl}) can serve as initial values for inflation. At first glance, such an idea of the beginning of inflation cannot contradict {\em the constraint} $V(\varphi)\sim 10^{-10}M_P^4$ {\em in the last stages of inflation}. However, the situation has changed dramatically over the past few years.

Data of 
recent cosmological observations \cite{Planck}-\cite{BICEP and Keck} favor inflationary models with plateau potentials, and with the height of the plateau $V_{pl}\sim 10^{-10}M_P^4$.
There exist a number of field theory models in which the plateau-like potentials arise due to the implementation of various original ideas, and these potentials satisfy the CMB constraints.
These include  
 the Starobinsky model \cite{Star},
the Goncharov-Linde model \cite{GL1}, the Higgs inflation models \cite{Bardeen}-\cite{Higgs infl 11}.
 Of particular interest are $\alpha$-attractor models, which were initiated by the pioneering work \cite{KL1} and which have been intensively studied in recent years. To date, there is  the broad class of the cosmological attractor models \cite{KL2}-\cite{KL9}, which generalize most of the previously proposed models with plateau potentials.

Despite such an impressive success of plateau-like models compared to all other models, a lively discussion 
 ensued, during which even the very idea of inflation has been called into question \cite{Stein}-\cite{Stein1}.
An obvious disadvantage of the models with plateau potentials mentioned above is the infinite length of the potential energy density plateau. In such a theory, all initial values of the homogeneous component of the scalar field $\varphi$ are equally probable. Therefore, an excessively long duration of inflation is possible.
The main problem formulated in paper\cite{Stein} is also related to  the infinite length of the potential energy density plateau. If the height of the plateau is $V_{pl}\sim 10^{-10}M_P^4$, then there is a huge range of possible values of the initial kinetic and gradient energy densities greater than $V_{pl}$, up to the Planck density. Therefore, in contrast to the understanding developed in  first models of chaotic inflation \cite{Linde 1983}, \cite{Linde 1985} of how initial conditions for inflation arise, there is no reason to believe that conditions (\ref{Cond for begin infl}) necessary for the onset of inflation are satisfied. A possible response to this challenge may be to modify the model in such a way that the potential has a plateau of finite length, after which, at very large $\varphi$, the potential rapidly increases. An example of this type of model is the "singular $\alpha$-attractor" model proposed by Linde in \cite{singular alpha}, in which the simplest $\alpha$-attractor potential takes an exponentially steep form for very large $\varphi $. This makes it possible to provide conditions for power-law inflation, which starts at the Planck density, and thus makes it possible to solve the problem of initial conditions in the spirit of \cite{Linde 1985}. 

\item

{\em The problem of  initial cosmological singularity in inflationary cosmology}

According to the BGV theorem\cite{BGV} which strengthens earlier proofs of singularity theorems \cite{BV1}, \cite{BV2},  in inflationary cosmology
  almost all past-directed timelike and null geodesics cannot be extended to the past beyond some boundary $\mathcal{B}$.
 The statement of the BGW theorem is quite general because it is based on a kinematic argument. The main problem that inevitably follows  from the statement of the BGV theorem is that the inflationary universe must have had some kind of beginning, and, therefore, some new physics is necessary in order
to determine the correct conditions at the boundary $\mathcal{B}$. However, the BGW theorem says nothing about the boundary conditions on 
$\mathcal{B}$, or even about its location.

\end{itemize}

It seems that the most attractive proposals for overcoming this problem are based on  the  idea of the possible existence of inflationary spacetime regions with opposite directions of the thermodynamic arrow of time \cite{termod arr1}-\cite{Vil arrows}. 
Taking this idea as a  hint and combining it with the existence of a spacelike boundary surface $\mathcal{B}$ predicted by the BGV theorem, in this paper we will try to build a model from the dynamics of which the following geometric approach to solving the problem will follow.

\begin{itemize}

\item

 Suppose that the boundary $\mathcal{B}$ is a spacelike hypersurface in the space-time manifold $M_4$ that separates submanifolds 
${M_4^{(+)}}$ and ${M_4^{(-)}}$ of  ${M_4}$ with opposite arrows of time, but with the same space-orientation. As a result, the  space-time orientations of the submanifolds ${M_4^{(+)}}$ and ${M_4^{(-)}}$ are opposite. Consequently,  the space-time manifold ${M_4}={M_4^{(+)}}\cup{M_4^{(-)}}$ turns out to be non-orientable. 
These conclusions are based on the definition, where the orientation is determined by the coordinate atlas (collection of maps). But n-dimensional differentiable manifolds (with or without boundary) allow one to define an orientation in an equivalent way using the n-form.
 Therefore, the orientations of the submanifolds $M_4^{(+)}$ and $M_4^{(-)}$ can be equivalently defined by either  oriented atlases or oriented 4-forms.
\footnote{More mathematical details will be presented in Sec.4.2.}

\end{itemize}

In the model studied in this paper, it is assumed that the space-time manifold $M_4$ is initially orientable. It will be shown that the geometric pattern described above is generated by the dynamics of the model, i.e., is the result of {\em spontaneous violation of the orientability of $M_4$.}
It turns out that this effect arises quite naturally in the Two-Measure Theory (TMT), where in the  integral of the primordial action, along with terms with the standard volume measure $dV_g=\sqrt{-g}d^4x$, there are terms with an alternative, metric independent volume measure $dV_{\Upsilon}$
 constructed as the following 4-form, globally defined on the orientable space-time manifold $M_4$ using 4 scalar functions $\varphi_a$ ($a =1 ,..,4$)
\begin{eqnarray}
dV_{\Upsilon}=\Upsilon d^4x\equiv
\varepsilon^{\mu\nu\gamma\beta}\varepsilon_{abcd}\partial_{\mu}\varphi_{a}
\partial_{\nu}\varphi_{b}\partial_{\gamma}\varphi_{c}
\partial_{\beta}\varphi_{d}d^4x
\nonumber\\
=4!d\varphi_1\wedge d\varphi_2\wedge d\varphi_3\wedge d\varphi_4.
\label{Phi}
\end{eqnarray}
Here $\Upsilon$ is a scalar density, that is under general coordinate transformations with positive Jacobian it has the same transformation law as $\sqrt{-g}$.
In contrast to the density $\sqrt{-g}$ of the standard volume measure $dV_g$, the density $\Upsilon$ of the 4-form $dV_{\Upsilon}$ is  sign-indefinite and,
due to its continuity, can also take zero values in the general case.
We will show that in the inflationary model studied in this paper, $\Upsilon$  can indeed take on a zero value  {\bf as a result of the dynamics} and, thus,  {\bf  the boundary $\mathcal{B}$  naturally arises  in the form of a spacelike hypersurface $\Upsilon(x)=0$.}.

At the end of this introductory part, it is necessary to briefly formulate the basics and main results of the TMT model under study, which demonstrate the close relationship between the problem of the initial cosmological singularity and the problem of initial conditions for inflation.

a) The primordial action\footnote{In "conventional" alternative theories of gravity, the original action differs from the Einstein-Hilbert action only in the form of the Lagrangian. In TMT, the main difference lies in the volume measure, although a modification of the Lagrangian is also possible. Therefore, the original action of TMT differs from the Einstein-Hilbert action in the form of the Lagrangian {\em density}. From this it follows that it would not be  quite correct to use the term Jordan frame for variables used in the original action of TMT. In order to emphasize this difference we will use the terms "primordial variables" and  "primordial action" instead of "variables in the original frame" and "original action".}
 of the TMT cosmological model is formulated  in the Palatini formalism\footnote{
 Ref. \cite{Demir} is an example for studying inflation with non-minimal coupling in the Palatini formulation. } 
on the  differentiable (and orientable) space-time manifold  ${M_4}$ where the following smooth variables are globally defined:  
four scalar functions $\varphi_a$ $(a=1,..,4)$,  affine connection $\Gamma^{\lambda}_{\mu\nu}$, primordial (original) metric tensor $g_{\mu\nu}$,
and the scalar field $\phi$ non-minimally coupled to gravity. The potential of the scalar field $\phi$ entering the  primordial  action has "the maximally possible standard form"
\begin{equation}
V(\phi)=\frac{1}{2}m^2\phi^2+\frac{\lambda}{4}\phi^4 \qquad \text{with} \qquad  m^2>0.
\label{V m2phi2}
\end{equation} 
The TMT effective potential of the canonically normalized scalar field $\varphi$ turns out to be of the shape which is a modification of the $T$-model type potential.

b) In a wide range of model parameters, a hypersurface $\Upsilon(x)=0$ arises that splits the manifold ${M_4}$ into two submanifolds ${M_4^{(+)}}$ and ${M_4^{(-)}}$ with opposite signs of $\Upsilon$ corresponding to the  opposite space-time orientations. This effect is an inevitable consequence of the dynamics of the model precisely due to the fact that, along with other variables, the action varies with respect to the functions $\varphi_a$ from which the 4-form $dV_{\Upsilon}=\Upsilon d^4x$ is built.

c) With an appropriate choice of an arbitrary integration constant, the universe, which has properties corresponding to ours, turns out to be in the submanifold ${M_4^{(+)}}$, where $\Upsilon>0$.
In our Universe, for admissible values of the canonically normalized scalar field $\varphi$, there is a limiting upper bound $\varphi_0$, which arises from the fact that $\Upsilon\rightarrow 0^+$ as $\varphi(x)\rightarrow\varphi_0^{\,\,\,-}$. It follows from this that in a homogeneous and isotropic universe the inflationary solution for the classical scalar field $\varphi(t)$ cannot be continued into the infinite past. The boundary $\mathcal{B}$ appearing in the BGV theorem arises as a hypersurface $\Upsilon(x)=0$ whose position is determined by the maximum allowable value $\varphi_0$ of the classical scalar field 
$\varphi$. Note that the model equations are not applicable at the boundary $\mathcal{B}$  itself, but the Hubble parameter remains finite at any proximity to the  boundary $\mathcal{B}$.

d) If in our Universe the initial kinetic energy density  $\rho_{kin}^{(in)}$ prevails over initial gradient energy density $\rho_{grad}^{(in)}$ then
there is an interval of initial values $\varphi_{in}^{(min)}\leq  \varphi_{in}
<\varphi_0$, where  $\rho_{kin}^{(in)}$ and $\rho_{grad}^{(in)}$ cannot exceed the potential energy density and hence  the initial conditions necessary for the beginning of inflation are satisfied.

e) Behind the boundary $\mathcal{B}$, that is in ${M_4^{(-)}}$, where $\Upsilon<0$, a variety of different universes with different vacuum states of the classical scalar field $\varphi$ can be realized by an appropriate choice  of the integration constant $\mathcal{M}$ and the value of ${\check\varphi}_0$ such that  $\Upsilon\rightarrow 0^-$ when $\varphi\rightarrow{\check\varphi}_0$. As a result, under almost all possible  left-handed boundary conditions on $\mathcal{B}$ the metric tensor in the Einstein frame has a jump discontinuity on $\mathcal{B}$, so the Christoffel connection coefficients are not defined on the spacelike hypersurface $\Upsilon=0$. Therefore, 
if $\varphi_{in}^{(min)}\leq \varphi_{in}<\varphi_0$ and $\rho_{kin}^{(in)}>\rho_{grad}^{(in)}$ , then there was an inflationary stage in the history of our Universe and the timelike geodesics cannot be extended to the past beyond the hypersurface $\Upsilon=0$.

f) The meaning of the title of the paper is precisely that the maximum allowable value $\varphi_0$ determines the position of the boundary $\mathcal{B}$ and is the exact upper limit of the interval of $\varphi$ in which the beginning of inflation is guaranteed.

The organization of paper is as follows.
In Sec.2, the simplest TMT model of the scalar field $\phi$ with potential (\ref{V m2phi2}) non-minimally coupled to gravity is formulated and studied. The use of the volume measure $\Upsilon d^4x$  in the primordial action
 naturally leads to the need, along with the usual vacuum-like term, to include in the primordial action a new kind of vacuum-like term with the corresponding model parameter $V_2$ of   dimensionality $(mass)^4$. It is shown that in the case of a zero effective cosmological constant, the TMT effective  potential of the canonically normalized scalar field $\varphi$ has a form similar to the potential of the $T$-model \cite{KL1}, but with a plateau of finite length: for $\varphi$ greater than some value $\varphi_*$, the potential becomes exponentially steep. The length of the plateau, and hence the duration of the quasi-de Sitter inflation, is controlled by the parameter $V_2$.
To avoid a misconception about the requirement that $\Upsilon$ be sign-definite, it is necessary to clearly and unambiguously formulate the following result of the model: {\bf the system of equations of motion obtained from the principle of least action has nontrivial solutions only if the sign of $\Upsilon$ does not change during the entire evolution of the Universe.}
 We show in Sec.2 that due to the sign-definiteness of $\Upsilon$ in our Universe, only the values $\varphi<\varphi_*$ are possible and limit
$\varphi\rightarrow \varphi_*$ corresponds to the process of approaching the boundary $\mathcal{B}$ whose equation  is 
\begin{equation}
\Upsilon(x)=0.
\label{ups 0}
\end{equation}
In Sec.3, we study a more general model that takes into account the arbitrariness in the choice of coefficients in a linear combination of volume elements
$dV_g=\sqrt{-g}d^4x$ and $dV_{\Upsilon}=\Upsilon d^4x$. It turns out that in such a model, the TMT effective potential has two plateaus. With an appropriate choice of parameters, for $\varphi<\varphi_0$ ($\varphi_0$ is close to $\varphi_*$), the shape of the potential is almost the same as in the model of Sec.2, but for  $\varphi>\varphi_0$, instead of an almost exponential growth, the potential has a second, higher plateau of infinite length.
In addition, in this model, when studying the possible initial conditions for inflation, it is found that the above requirement that $\Upsilon$ be sign-definite imposes restrictions on the admissible values of the kinetic and gradient energy densities. A detailed analysis given in  Appendix B shows that there is an  interval of initial values $\varphi_{in}$, bounded from above by $\varphi_0$, in which the initial conditions necessary for inflation are guaranteed to be satisfied; this requires the only additional condition, which is that  the initial kinetic energy density is greater than the initial gradient energy density. In Sec.4 we show that the boundary hypersurface $\Upsilon(x)=0$ is spacelike. Analysing the  one-sided boundary conditions on $\Upsilon(x)=0$ we find that the metric tensor in the Einstein frame has discontinuety on $\Upsilon(x)=0$, from which it follows that past directed geodesics cannot be continued beyond the boundary $\mathcal{B}$. After this we shortly review  some mathematical aspects concerning orientability that allows us to interpret the obtained effect as a spontaneous violation of the orientability of the spacetime manifold $M_4$.
Sec.5 is devoted to a discussion of the radical changes introduced by TMT into field theory, due to which the results of this work were achieved.
To facilitate understanding of the paper for the reader not familiar with TMT, 
 in Appendix A we give  a very brief introduction to the basics of TMT and 
 describe the procedure of TMT, the implementation of which allows one to obtain  the equations, the potential of the scalar field and the action in the Einstein frame, which are further called TMT effective equations, the TMT effective potential and the TMT effective action, respectively.
Appendix C outlines the algorithm that can be used to calculate quantum corrections to the model studied in this paper.

\section{A simplest field theory model}

\subsection{The primordial action and equations of motion in the Einstein frame}

Consider a model that includes gravity,  the inflaton field with the canonical kinetic term, a nonminimal inflaton-to-scalar curvature coupling and vacuum-like terms. 
 In this section, the primordial action is chosen as follows
\begin{equation}
S=\int d^4x(\sqrt{-g}+\Upsilon)\left[L_{gr}+L_{\phi}+L_{nonmin}\right]+S_{vac},
\label{S-gr-infl}
\end{equation} 
where 
\begin{equation}
 L_{gr}=-\frac{M_P^2}{2}R(\Gamma,g); \qquad  L_{\phi}= \frac{1}{2}g^{\alpha\beta}\phi_{,\alpha}\phi_{,\beta}-V(\phi); \qquad
 L_{nonmin}=-\frac{1}{2}\xi R(\Gamma,g)\phi^2
\label{3 Lagr}
\end{equation} 
Here $M_P$  is the reduced Planck mass;  \,  $\Gamma$ stands for affine connection; $R(\Gamma,g)=g^{\mu\nu}R_{\mu\nu}(\Gamma)$,
$R_{\mu\nu}(\Gamma)=R^{\lambda}_{\mu\nu\lambda}(\Gamma)$ and
$R^{\lambda}_{\mu\nu\sigma}(\Gamma)\equiv \Gamma^{\lambda}_{\mu\nu
,\sigma}+ \Gamma^{\lambda}_{\gamma\sigma}\Gamma^{\gamma}_{\mu\nu}-
(\nu\leftrightarrow\sigma)$. In our notations  the  parameter $\xi$ of  non-minimal coupling of a massless scalar field in the case of a conformal coupling in a theory with only the volume element $\sqrt{-g}d^4x$ would be equal to $\xi=-\frac{1}{6}$.

Contribution of the  vacuum-like terms to the  primordial action is defined by
\begin{equation}
S_{vac}=\int d^4x\left(-\sqrt{-g}V_1-\frac{\Upsilon^2}{\sqrt{-g}}V_2\right).
\label{S_vac}
\end{equation} 
If the first term in Eq.(\ref{S_vac})   was present in Einstein's GR,  
$V_1$ would be a cosmological constant. 
The term with $V_2$ was first introduced in Ref. \cite{GK2}. The first reason for adding the term with $V_2$ is that the term
 $\propto \Upsilon=\zeta(x)\sqrt{-g}$, which can be expected as the contribution of quantum-gravitational effects
 to a vacuum-like action, does not contribute to the equations of motion. Therefore, the next  in powers of $\zeta$ 
vacuum-like term can be of the form $\propto \zeta^2\sqrt{-g}= \Upsilon^2/\sqrt{-g}$. 
There are also  more pragmatic  reasons. It turns out that thanks to this term,  the corresponding TMT effective potential  and the TMT effective action  acquire fundamentally new important properties without which the goals of this paper could not be achieved.

In general, there is no reason why the coefficients in linear combinations of $\Upsilon$ and $\sqrt{-g}$ in volume elements of different terms in the action should be the same. In the gravitational term with $L_{gr}$, the common factor of  the linear combination can be absorbed by redefining Newton's constant. After that, by rescaling the fields $\varphi_a$ in Eq.({\ref{Phi}}), the volume element in the gravitational term with $L_{gr}$ becomes as in Eq.(\ref{S-gr-infl}). Then in all other contributions to the  primordial action there is only the freedom to absorb the common factor by rescaling $\phi$, the parameters in the  primordial potential $V(\phi)$ and $\xi$. As a result, the corresponding volume elements can have the form $\left(b_i\sqrt{-g}+s\Upsilon\right)d^4x$, where $b_i$ and  $s=\pm 1$ are arbitrary model parameters.  The appearance of the parameter  $s$ is due to the fact that in the  primordial action $\Upsilon$ can equally be both positive and negative. Therefore, in general, $\Upsilon$ can  enter the primordial action with different signs. In this paper, we will restrict ourselves to the case $s=1$.
The choice of the same volume element $(\sqrt{-g}+\Upsilon)d^4x$ for all terms in Eq.(\ref{S-gr-infl}) made in this section means that we are dealing with
{\em  the simplest version of the model}.

For the  primordial inflaton potential $V(\phi)$ we choose a simple model of massive scalar field with quartic selfinteraction, Eq.(\ref{V m2phi2}).

Now, following the prescription of the TMT procedure described in Appendix A, we consider the equations of motion following from the  primordial action  (\ref{S-gr-infl}).
Varying the action with respect to
 scalar functions $\varphi_{a}$ of which $\Upsilon$ is built we get 
\begin{equation}
B^{\mu}_{a}\partial_{\mu}\left[L_{gr}+L_{\phi}+L_{nonmin}-2\zeta V_2\right]=0 \quad \text{where} \quad
B^{\mu}_{a}=\varepsilon^{\mu\nu\alpha\beta}\varepsilon_{abcd}
\partial_{\nu}\varphi_{b}\partial_{\alpha}\varphi_{c}
\partial_{\beta}\varphi_{d},
\label{varphiB}
\end{equation}
and where the following scalar $\zeta(x)$ appears:
\begin{equation}
 \zeta(x)\stackrel{\mathrm{def}}{=} \frac{dV_{\Upsilon}}{dV_g}\equiv\frac{\Upsilon}{\sqrt{-g}}   
 \label{zeta}
\end{equation}

Since $Det (B^{\mu}_{a}) = \frac{4^{-4}}{4!}\Upsilon^{3}$ it follows
that if
 \begin{equation}
\quad \textsf{everywhere} \quad \Upsilon(x)\neq 0, 
\label{Phi neq 0}
\end{equation}
the equality
\begin{equation}
-\frac{M_P^2}{2}\left(1+\xi \frac{\phi^2}{M_P^2}\right)R(\Gamma,g)+\frac{1}{2}g^{\alpha\beta}\phi_{,\alpha}\phi_{,\beta}
-V(\phi)-2\zeta V_2={\mathcal M}
\label{var varphi}
\end{equation} 
must be satisfied, where  ${\mathcal M}$ is a constant of
integration with the dimension of $(mass)^4$.

Variation with respect to $g^{\mu\nu}$ yields  the equation
\begin{eqnarray}
&&(1+\zeta)\left[-\frac{M_P^2}{2}\left(1+\xi \frac{\phi^2}{M_P^2}\right)R_{\mu\nu}(\Gamma)+\frac{1}{2}\phi_{,\mu}\phi_{,\nu}\right]
\nonumber
\\
&-& \frac{1}{2}g_{\mu\nu}\left[-\frac{M_P^2}{2}\left(1+\xi\frac{\phi^2}{M_P^2}\right)R(\Gamma, g) 
 +\frac{1}{2}g^{\alpha\beta}\phi_{,\alpha}\phi_{,\beta}-V(\phi) -V_1+ \zeta^2 V_2 \right]=0.
 \label{Grav.eq}
\end{eqnarray}
The trace of Eq.(\ref{Grav.eq}) is the following
\begin{equation}
(\zeta -1)\left[-\frac{M_P^2}{2}\left(1+\xi\frac{\phi^2}{M_P^2}\right)R(\Gamma,g)+\frac{1}{2}g^{\alpha\beta}\phi_{,\alpha}\phi_{,\beta}\right]
+2V(\phi)+2V_1-2\zeta^2 V_2=0.
\label{trace of grav}
\end{equation} 
Eliminating  the expression  $-\frac{1}{2}M_P^2\left(1+\xi\frac{\phi^2}{M_P^2}\right) R(\Gamma, g)$ from Eqs.(\ref{var varphi}) and  
(\ref{trace of grav})
 one can see that the term  $2\zeta^2V_2$ is canceled, and the consistency of these equations requires that the scalar function $\zeta(x)$ satisfies the following relation
\begin{equation}
\zeta=\zeta(\phi(x))=
\frac{{\mathcal M}-2V_1-V(\phi)} {{\mathcal M}-2V_2+V(\phi)}
 \label{Constraint}
\end{equation}
describing $\zeta(x)$ as the local function of the inflaton $\phi(x)$ \footnote{Note that the model should not allow the possibility for $\zeta$ to cross the value -1, since this would mean a change in the sign of the volume measure in the primordial action (\ref{S-gr-infl}). Using the form of $V(\phi)$, Eq.(\ref{V m2phi2}), we see that $\zeta(\phi)\to -1$ as $\phi\to \infty$.}. 
Following the terminology of earlier works \cite{GK2}-\cite{GK6}, we will call this a constraint.
It should be noted that in the Palatini formulation $\zeta(x)$ is not a physical degree of freedom. Therefore, when we call Eq.(\ref{Constraint})  a constraint, we must keep in mind that it is different in meaning from the usual constraint in the field theory models, where it describes the relationship between dynamical degrees of freedom.

The inflaton  equation reads
\begin{equation}
\frac{1}{\sqrt{-g}}\partial_{\mu}\left[(1+\zeta)\sqrt{-g}g^{\mu\nu}\partial_{\nu}\phi\right]
+(1+\zeta)\left[V^{\prime}(\phi)+\xi R(\Gamma, g)\phi\right]=0.
 \label{phi-original}
\end{equation}

Variation of the affine connection yields the equation that has been   solved in Ref. \cite{GK2}
 for the simpler case. For the case of the action (\ref{S-gr-infl}) containing a non-minimal coupling  the result is
\begin{equation}
\Gamma^{\lambda}_{\mu\nu}=\{^{\lambda}_{\mu\nu}\}+ (\delta^{\lambda}_{\mu}\sigma,_{\nu}
+\delta^{\lambda}_{\nu}\sigma,_{\mu}-
\sigma,_{\beta}g_{\mu\nu}g^{\lambda\beta}),
 \label{GAM2}
\end{equation}
where $\{^{\lambda}_{\mu\nu}\}$  are the Christoffel's connection
coefficients of the metric $g_{\mu\nu}$ and
\begin{equation}
\sigma(x)=\frac{1}{2}\ln\left[\Bigl(1+\zeta(\phi(x))\Bigr)\left(1+\xi\frac{\phi^2(x)}{M_P^2}\right)\right].
\label{ln in Gamma}
\end{equation}
If $\sigma(x)\neq const.$ the metricity condition does not hold and consequently
geometry of the space-time with the metric $g_{\mu\nu}$ is generically non-Riemannian. In this paper, we will totally ignore a possibility to incorporate the torsion tensor, which could be an additional source for the space-time  to be different from Riemannian. 

It is easy to see that  the transformation of the metric
\begin{equation}
\tilde{g}_{\mu\nu}=(1+\zeta(\phi))\left(1+\xi\frac{\phi^2(x)}{M_P^2}\right)g_{\mu\nu}
 \label{gmunuEin}
\end{equation}
turns  the connection $\Gamma^{\lambda}_{\mu\nu}$ into
the Christoffel connection coefficients of the metric
$\tilde{g}_{\mu\nu}$ and the space-time turns into (pseudo)
Riemannian. 
Gravitational equations
(\ref{Grav.eq}) expressed in terms of the metric $\tilde{g}_{\mu\nu}$ take the
canonical GR  form 
\begin{equation}
R_{\mu\nu}(\tilde{g})-\frac{1}{2}\tilde{g}_{\mu\nu}R(\tilde{g})=\frac{1}{M_P^2}T_{\mu\nu}^{(eff)}
\label{grav eq Ein}
\end{equation}
with the same Newton constant as in the original frame.
Here $R_{\mu\nu}(\tilde{g})$ and $R(\tilde{g})$ are the Ricci tensor and the scalar curvature of the metric $\tilde{g}_{\mu\nu}$, respectively.
 Therefore the set of dynamical variables using the metric $\tilde{g}_{\mu\nu}$ can be called the Einstein frame. 
$T_{\mu\nu}^{(eff)}$ on the right side of the Einstein equations (\ref{grav eq Ein})  is {\em the  TMT effective energy-momentum tensor}, the occurrence of which is described in Appendix A  as step 6 in the TMT procedure. In the model under study, $T_{\mu\nu}^{(eff)}$ has the form
\begin{equation}
T_{\mu\nu}^{(eff)}=\frac{1}{1+\xi\frac{\phi^2}{M_P^2}}\left(\phi_{,\mu}\phi_{,\nu}
- \tilde{g}_{\mu\nu}X_{\phi}\right)
+\tilde{g}_{\mu\nu}U_{eff}(\phi,\zeta(\phi);{\mathcal M}), \, \text{where} \,\,  X_{\phi}=\frac{1}{2}\tilde{g}^{\alpha\beta}\phi_{,\alpha}\phi_{,\beta};
 \label{Tmn}
\end{equation}
 {\em the TMT effective potential} $U_{eff}(\phi,\zeta(\phi);{\mathcal M})$ appears as the following function of $\phi$ and the integration constant ${\mathcal M}$
\begin{equation}
U_{eff}(\phi,\zeta(\phi);{\mathcal M})=\frac{1}{\left(1+\xi\frac{\phi^2}{M_P^2}\right)^2}\left[V_2+\frac{{\mathcal M}-V_1-V_2}{[1+\zeta(\phi)]^2}\right],
 \label{u zeta}
\end{equation}
where  $\zeta(\phi)$ is determined by  the constraint (\ref{Constraint}). 

The role of the non-minimal coupling in the flattening of the inflationary potential and in the attractor for inflation at strong coupling  is well known \cite{Bardeen}-\cite{Higgs infl 11}, \cite{KL non-minimal and attrtact 1}, \cite{KL non-minimal and attrtact 2}.
It is noteworthy that in addition to the factor $\left(1+\xi\frac{\phi^2}{M_P^2}\right)^{-2}$ generated by a non-minimal coupling, $U(\phi,\zeta(\phi);{\mathcal M})$ contains also the factor $(1+\zeta(\phi))^{-2}$, the occurrence of which is due to the inherent properties of TMT. 
As we shall  see, {\em the fact that the  plateau of the TMT effective potential has a finite length is one of the most important effects of the scalar $\zeta(\phi)$}.

If  ${\mathcal M}\neq V_1+V_2$, after making use the constraint (\ref{Constraint}) the TMT effective potential $U_{eff}(\phi,\zeta(\phi);{\mathcal M})$ can be represented in the $\zeta$-independent form 
\begin{equation}
U_{eff}(\phi;{\mathcal M})=\frac{1}{\left(1+\xi\frac{\phi^2}{M_P^2}\right)^2}\left[V_2+\frac{\left[{\mathcal M}-2V_2+V(\phi)\right]^2}{4({\mathcal M}-V_1-V_2)}\right].
\label{u zeta independent}
\end{equation}
Recall that $V(\phi)$ is defined by Eq.(\ref{V m2phi2}).
The presence of the factor $(1+\zeta)^{-2}$ in $U_{eff}(\phi,\zeta(\phi);{\mathcal M})$ causes $V^2 (\phi)/\left(1+\xi\phi^2/M_P^2\right)^2$ to appear in $U_{eff}(\phi;{\mathcal M})$,  which for sufficiently large $\phi$ leads to a change in the shape of the TMT effective potential from plateau to steep rise (see also footnote 4).

For further study, it is convenient to rewrite $U_{eff}(\phi;{\mathcal M})$ by extracting a term equal to the value of $U_{eff}(\phi;{\mathcal M})$ at $\phi=0$
\begin{equation}
\Lambda({\mathcal M})=U_{eff}(\phi=0;{\mathcal M})=
 \frac{{\mathcal M}^2-4V_1V_2}{4({\mathcal M}-V_1-V_2)}.
 \label{L of M}
\end{equation}
Then we get
\begin{equation}
U_{eff}(\phi;{\mathcal M})=\Lambda({\mathcal M})+V_{eff}(\phi;{\mathcal M}),
 \label{Lambda + V phi}
\end{equation}
where 
\begin{equation}
V_{eff}(\phi;{\mathcal M})=\frac{2({\mathcal M}-2V_2)V(\phi)+\left[V(\phi)\right]^2}{4({\mathcal M}-V_1-V_2)\left(1+\xi\frac{ \phi^2}{M_P^2}\right)^2}
-\Lambda({\mathcal M})\left(1-\frac{1}{\left(1+\xi\frac{ \phi^2}{M_P^2}\right)^2}\right).
\label{u eff via Vphi}
\end{equation}

After passing to the Einstein frame in the inflaton equation (\ref{phi-original}) one must substitute the expression for the scalar curvature derived from the Einstein equations (\ref{grav eq Ein}). And finally, using the constraint (\ref{Constraint}), we get  the inflaton equation  in the following form
\begin{eqnarray}
&&\frac{1}{\sqrt{\tilde{g}}}\partial_{\mu}\left(\frac{1}{1+\xi\frac{\phi^2}{M_P^2}}\sqrt{\tilde{g}}\tilde{g}^{\mu\nu}\partial_{\nu}\phi\right)+
\frac{\xi\phi}{M_P^2\left(1+\xi\frac{\phi^2}{M_P^2}\right)^2}\tilde{g}^{\alpha\beta}\partial_{\alpha}\phi\partial_{\beta}\phi
\label{phi eq Ein without zeta}
\\
&+&
\frac{{\mathcal M}-2V_2+V(\phi)}{2({\mathcal M}-V_1-V_2)\left(1+\xi\frac{\phi^2}{M_P^2}\right)^2}V^{\prime}(\phi)-
\frac{\xi\phi}{M_P^2\left(1+\xi\frac{\phi^2}{M_P^2}\right)^3}
\left[\frac{\left[{\mathcal M}-2V_2+V(\phi)\right]^2}{{\mathcal M}-V_1-V_2}+4V_2\right]=0.
\nonumber
\end{eqnarray}

\subsection{The zero cosmological constant case}

The arbitrariness in the value of the integration constant ${\mathcal M}$ allows one to study cosmological models with an arbitrary cosmological constant. In this paper, we will restrict ourselves to a model with zero vacuum energy density, which is a fairly good approximation when studying the inflationary epoch. If $\phi=0$ is the position of the global minimum of the TMT effective potential $U_{eff}(\phi;{\mathcal M})$ described by 
Eqs.(\ref{u zeta independent})-(\ref{u eff via Vphi}), i.e. $\phi=0$ is the vacuum state of the classical scalar field $\phi(x)$, then $\Lambda({\mathcal M})$ is the cosmological constant. 
As can be seen from Eqs.(\ref{Constraint}) and (\ref{V m2phi2}), in the vacuum the expression for the scalar $\zeta$ has the form
\begin{equation}
\zeta_v=
\frac{{\mathcal M}-2V_1}{{\mathcal M}-2V_2}.
\label{zeta in vac}
\end{equation}
The zero value of the cosmological constant is reached if the integration constant ${\mathcal M}$ satisfies the relation
\begin{equation}
{\mathcal M}_0^2=4V_1V_2,
\label{M Lambda 0}
\end{equation}
where the subscript $0$ indicates that $\Lambda({\mathcal M}_0)=0$. To ensure the possibility of $\Lambda({\mathcal M}_0)=0$, the model parameters $V_1$ and $V_2$ must have the same sign. 
It can be shown by direct detailed verification that results of interest are obtained if the parameters $V_1$, $V_2$  and the integration constant 
${\mathcal M}_0$ are chosen so that
\begin{equation}
V_1<0; \quad V_2<0; \quad {\mathcal M}_0=2\sqrt{V_1V_2}.
\label{V1 V2 M}
\end{equation}
 Then it follows from  Eq.(\ref{zeta in vac}) that the value of $\zeta$ in the vacuum with zero energy density is
\begin{equation}
\zeta_v=\sqrt{\frac{V_1}{V_2}}>0.
\label{zeta in vac 0}
\end{equation}

In the chosen case of a zero cosmological constant,  the TMT effective potential $U_{eff}(\phi;{\mathcal M})$ defined by Eqs.(\ref{u zeta independent})-(\ref{u eff via Vphi})  reduces to $V_{eff}^{(0)}(\phi) \stackrel{\mathrm{def}}{=}
U_{eff}(\phi;{\mathcal M}_0)=V_{eff}( \phi;\mathcal{ M }_0)$, which has the form
\begin{equation}
V_{eff}^{(0)}(\phi) =\frac{1}{(1+\zeta_v)\left(1+\xi \frac{\phi^2}{M_P^2}\right)^2}
\left[V(\phi)+ 
\frac{\left(V(\phi)\right)^2}{4|V_2|(1+\zeta_v)}\right]
\label{V eff Lambda 0 without form of V}
\end{equation}
Accordingly, when choosing (\ref{V1 V2 M}), the inflaton equation (\ref{phi eq Ein without zeta}) reduces to
\begin{eqnarray}
&&\frac{1}{\sqrt{\tilde{g}}}\partial_{\mu}\left(\frac{1}{1+\xi\frac{\phi^2}{M_P^2}}\sqrt{\tilde{g}}\tilde{g}^{\mu\nu}\partial_{\nu}\phi\right)+
\frac{\xi\phi}{M_P^2\left(1+\xi\frac{\phi^2}{M_P^2}\right)^2}\tilde{g}^{\alpha\beta}\partial_{\alpha}\phi\partial_{\beta}\phi
\nonumber
\\
&+&
\frac{2|V_2|(1+\zeta_v)+V(\phi)}{2|V_2|(1+\zeta_v)^2\left(1+\xi\frac{\phi^2}{M_P^2}\right)^2}V^{\prime}(\phi)-
\frac{\xi\phi \left(V(\phi)\right)^2}{|V_2|M_P^2(1+\zeta_v)^2\left(1+\xi\frac{\phi^2}{M_P^2}\right)^3}
=0.
 \label{phi eq Ein with zero Lambda}
\end{eqnarray}

Following the TMT procedure, we are left with the 7th step.
Namely, it is necessary to make sure that Eq.(\ref{phi eq Ein with zero Lambda}) for the inflaton field and the Einstein equations (\ref{grav eq Ein}) with the energy-momentum tensor $T_{\mu\nu}^{(eff)}$ described by Eqs.(\ref{Tmn})-(\ref{u zeta independent}) with $\Lambda({\mathcal M}_0)=0$ and (\ref{V eff Lambda 0 without form of V}) are self-consistent. This  can be done if these equations can be obtained from some effective action.
As usual, if $\tilde{g}^{\alpha\beta}\phi_{,\alpha}\phi_{,\beta}>0$,  the energy-momentum tensor $T_{\mu\nu}^{(eff)}$ can be rewritten in the form of a perfect fluid. Then the pressure density plays the role of the Lagrangian in the effective action. Thus we come to the following TMT effective action
\begin{equation}
S_{eff}=\int \left(-\frac{M_P^2}{2}R(\tilde{g})+\frac{1}{2\left(1+\xi \frac{\phi^2}{M_P^2}\right)}\tilde{g}^{\alpha\beta}\phi_{,\alpha}\phi_{,\beta}-V_{eff}^{(0)}(\phi))\right)\sqrt{-\tilde{g}}d^4x
\label{Seff}
\end{equation}
It can be checked by direct calculations that the variation of $S_{eff}$ with respect to the fields $\tilde{g}^{\mu\nu}$ and $\phi$ actually leads to 
Eqs.(\ref{phi eq Ein with zero Lambda}) and (\ref{grav eq Ein}) with  $T_{\mu\nu}^{(eff)}$ described by Eqs.(\ref{Tmn})-(\ref{u zeta independent}) and (\ref{V eff Lambda 0 without form of V}) (with $\Lambda=0$).

To compare the predictions of the model under study with numerous models developed to describe inflationary cosmology, it is necessary to pass to the canonically normalized scalar field.  It turns out that the most interesting results are obtained if the non-minimal coupling constant $\xi$ is positive. Therefore, in addition to choosing negative values of the model parameters $V_1$ and $V_2$ and the integration constant $\mathcal{ M }_0=2\sqrt{V_1V_2}$, (see Eq.(\ref{V1 V2 M})), we will study the results of the model obtained for $\xi>0$.
Then the canonically normalized scalar field $\varphi$ can be easily found by solving the equation $\frac{d\phi}{d\varphi}=\sqrt{1+\xi\frac{\phi^2}{M_P^2}}$, which gives
\begin{equation}
\frac{\phi}{M_P}=\frac{1}{\sqrt{\xi}}{\sinh}\left(\sqrt{\xi}\frac{\varphi}{M_P}\right).
\label{phi via canonical varphi}
\end{equation}

The action (\ref{Seff}) is then given by
\begin{equation}
S_{eff}=\int \left(-\frac{M_P^2}{2}R(\tilde{g})+\frac{1}{2}\tilde{g}^{\alpha\beta}\varphi_{,\alpha}\varphi_{,\beta}-V_{eff}^{(0)}(\phi(\varphi))\right)\sqrt{-\tilde{g}}d^4x, 
\label{Seff varphi}
\end{equation}
where, after inserting the primordial potential (\ref{V m2phi2}), the TMT effective  potential expressed in terms of the canonically normalized field $\varphi$, $V_{eff}^{(0)}(\varphi)=V_{eff}^{(0)} (\phi(\varphi))$, has the form
\begin{eqnarray}
\begin{split}
V_{eff}^{(0)}(\varphi) =
\frac{M_P^4}{(1+\zeta_v)\xi}{\tanh}^4z
&
\left[\frac{\lambda}{4\xi}+
\frac{m^4}{16\xi|V_2|(1+\zeta_v)}+\frac{m^2}{2M_P^2\cdot {\sinh^2}z}
\right.
\\
&+
\left.
\frac{\lambda m^2M_P^2}{16\xi^2|V_2|(1+\zeta_v)}{\sinh}^2z+\frac{\lambda^2M_P^4}{64\xi^3|V_2|(1+\zeta_v)}{\sinh}^4z
\right],
\end{split}
\label{Veff varphi tanh}
\end{eqnarray}
where $z=\sqrt{\xi}\frac{\varphi}{M_P}$.

In the next subsection, we will see the need to compare the model under study with  the $\alpha$-atttactor models\cite{KL1}-\cite{KL9}, \cite{singular alpha}. Here it is worth paying attention to the difference between the change of variables $\phi\rightarrow\varphi$ described by Eq.(\ref{phi via canonical varphi}) and the change $\phi\rightarrow\varphi$
in the $\alpha$-atttactor models. In this regard, the models are fundamentally different. In $\alpha$-atttactor models, the kinetic term of the non-canonical scalar field $\phi$ has a pole at  the boundary of the moduli space. 
On the contrary, in the model under study, when $\xi>0$ is chosen, the noncanonical scalar field $\phi$ in the TMT effective  action (\ref{Seff}) does not contain a pole. This difference is insignificant in the limit of $\varphi\rightarrow 0$: in both models $\phi\rightarrow 0$ as 
$\varphi\rightarrow 0$. But for $\varphi\rightarrow \infty$, the corresponding behavior of $\phi$ in these models is completely different. In the TMT model under study, $\phi$ also tends to infinity. In $\alpha$-atttactor models, the canonical scalar field tends to infinity when the original non-canonical scalar field tends to the boundary of the moduli space.

In the zero cosmological constant case, the constraint (\ref{Constraint}) presented in terms of the canonically normalized field $\varphi$ takes the form
\begin{equation}
\zeta=\frac{2|V_2|\zeta_v(1+\zeta_v)-\frac{m^2M_P^2}{2\xi}\sinh^2z-\frac{\lambda}{4\xi^2}M_P^4\sinh^4z}
{2|V_2|(1+\zeta_v)+\frac{m^2M_P^2}{2\xi}\sinh^2z+\frac{\lambda}{4\xi^2}M_P^4\sinh^4z},
\quad \text{where} \quad z=\sqrt{\xi}\frac{\varphi}{M_P}.
\label{zeta fine tun}
\end{equation}

\subsection{Model parameters and preliminary discussion of the obtained modification of the T-model potential}

The model contains 5 parameters: $m^2>0$, $\lambda>0$, $\xi>0$, $V_1<0$ and $V_2<0$. This allows us to hope that by fitting the parameters it will be possible to obtain agreement between the predictions of the model   and the existing observational data (and, possibly, future ones). 
But in this paper we will restrict ourselves to studying the possibilities of solving the problem of the initial conditions for inflation and its connection with the problem of the initial cosmological singularity.

As we have seen, if $\phi=0$ is a vacuum state with zero energy density, then, instead of the parameter $V_1$, it is convenient to use the parameter $\zeta_v$ defined by Eq.(\ref{zeta in vac 0}). 
As noted above, the canonical variable $\varphi=0$ when $\phi=0$.
Expanding the TMT effective  potential (\ref{Veff varphi tanh}) near $\varphi=0$ we obtain the following expression for the square of the inflaton effective  mass
\begin{equation}
m^2_{eff}=\frac{m^2}{1+\zeta_v}.
\label{m2 eff}
\end{equation}

Assuming that the parameter $|V_1|$ does not exceed the parameter $|V_2|$, we obtain from Eq.(\ref{zeta in vac 0})
\begin{equation}
0<\zeta_v\leq 1
\label{m2 eff}
\end{equation}
As a numerical estimate, let us take the mass of the inflaton $m_{eff}\approx 2\cdot 10^{13}GeV$.  
Then for the mass parameter $m$ we have
\begin{equation}
m\sim 10^{13}GeV. 
\label{m estimate}
\end{equation}

The first two terms in square brackets in (\ref{Veff varphi tanh}) are responsible for  the start of the plateau when $\frac{\varphi}{M_P}\geq \frac{1}{\sqrt{\xi}}$.
The last two terms  in (\ref{Veff varphi tanh}) are responsible for the end of the plateau and the beginning of the almost exponential growth of 
$V_{eff}^{(0)}(\varphi)$. By adjusting the model parameters, one can move the  value of $\varphi$ at which the latter occurs to provide constraints on the initial conditions for a quasi-dS inflation.

We notice  that by choosing the value of the non-minimal coupling parameter $\xi=\frac{1}{6}$ (opposite in sign to the parameter of the conformal coupling ), we obtain an effective potential constructed from the hyperbolic functions like ${\tanh}\frac{\varphi}{\sqrt{6}M_P}$, depending on the same combination $\frac{\varphi}{\sqrt{6}M_P}$ as in the simplest T-Model obtained in the conformal theory \cite{KL1}. A more general case of  
the single field $\alpha$ attractor models \cite{KL3}, where   the hyperbolic functions depend on the  combination $\frac{\varphi}{\sqrt{6\alpha}M_P}$, corresponds to the choice $\xi=\frac{1}{6\alpha}$ in the model of this paper.

For the plateau height of $V_{eff}^{(0)}(\varphi)$ to satisfy the constraint on the inflationary energy scale \cite{Planck}, the choice of model parameters must ensure that
\begin{equation}
\frac{1}{4(1+\zeta_v)\xi^2}
\left[\lambda+
\frac{m^4}{4|V_2|(1+\zeta_v)}\right]\sim  10^{-10}.
\label{bound on parameters for hight}
\end{equation}
For further evaluation, we must consider the possible range of the  parameters $|V_2|$ and $\lambda$.
It is natural to assume that the vacuum-like parameters $V_1$ and $V_2$ can be of the order close or at least not much less than $M_P^4$.
Using (\ref{m estimate})  one can conclude that 
\begin{itemize}

\item

if $\lambda<\frac{m^4}{4(1+\zeta_v)|V_2|}$ then (\ref{bound on parameters for hight}) can be  satisfied if $|V_2|\lesssim(\frac{1}{2}\cdot 10^{16}GeV)^4$;

\item

if we prefer $|V_2|$ to be closer to the Planck scale, i.e. $|V_2|>(10^{16}GeV)^4$, then there must be
\begin{equation}
\frac{\lambda}{4\xi^2(1+\zeta_v)}\sim 10^{-10}, \quad \text{that is} \quad \lambda\sim\frac{1+\zeta_v}{\alpha^2} \cdot 10^{-11}.  
\label{lambda estim}
\end{equation}
In what follows, the last choice for $\lambda$ will be assumed.
\end{itemize}

\begin{figure}
\includegraphics[width=13.0cm,height=8cm]{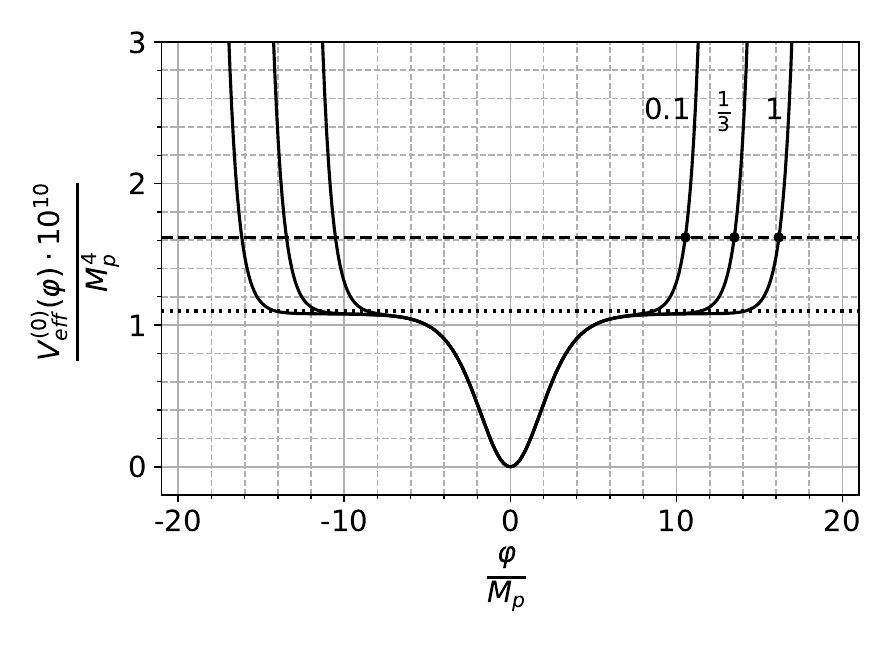}
\caption{Plots of the effective TMT potential $V_{eff}^{(0)}(\varphi)$ defined by Eq.(\ref{Veff varphi tanh}) show that the plateau length is  controlled by the parameter $|V_2|=(qM_P)^4$.
 As an example, three values  $q=1$, \, $q=\frac{1}{3}$, \, $q=0.1$ are chosen; each curve is labeled with the corresponding value of $q$.
The other parameters are the same for all curves: $\xi=\frac{1}{6}$, \, $\zeta_v=1$, \, $\lambda = 2.4\cdot 10^{-11}$, \, $m=1.9\cdot 10^{13}GeV$.  
The  points of intersection of the dashed line with the function curves correspond to $\varphi=\varphi_*$ defined by Eq.(\ref{condition zeta 0 fine tuned case}). $q$-dependent quantities $\varphi_*$ determine the maximum possible duration of inflation. But the values of $V_{eff}^{(0)}(\varphi_*)$ do not depend on $q$, and for the chosen parameters $V_{eff}^{(0)}(\varphi_*)$ exceeds the plateau height by approximately 1.5 times.
The latter means that at the very beginning, inflation can be driven by a scalar field with an almost exponentially growing potential.}
\label{fig1}
\end{figure}

Now there is a need to discuss the parameters of the model in more detail.
Note that with the chosen parameters $m$ and $\lambda$, the height of the potential energy plateau of the TMT effective potential $V_{eff}^{(0)}(\varphi)$, Eq.(\ref{Veff varphi tanh}), is controlled mainly by $\lambda$, and more precise tuning can be performed using $\xi$ and $\zeta_v$. The set of these four parameters, together with $|V_2|$ discussed below, provides a wide range of possibilities for adjusting the shape of  $V_{eff}^{(0)}(\varphi)$ to the constraints imposed by current and future cosmological observations. 
 By the number of model parameters, the effective TMT potential $V_{eff}^{(0)}(\varphi)$ is very different from the potential in  the $\alpha$-attractor models\cite{KL3}, where only two parameters ($m$ and $\alpha$) are present.  However, this is not the only difference.  Due to the presence of an almost exponentially growing "tail", $V_{eff}^{(0)}(\varphi)$ should be compared with the potential proposed by Linde  in Ref.\cite{singular alpha}  as an example of a  singular $\alpha$-attractor model   to solve  the problem of initial conditions for inflation.

After we have chosen estimates for the parameters  $m$,  $\lambda$,  $\xi$, $\zeta_v$,
the only remaining free  parameter is $V_2<0$. It is convenient to use the parametrization  $|V_2|=(q\cdot M_P)^4$. In order to get an idea of the effect of a parameter $|V_2|$, let us consider the following three cases: 1)  $q=1$, which corresponds to  $|V_2|=M_P^4\approx (2.44\cdot 10^{18}GeV)^4$; \, 2) $q=\frac{1}{3}$, which corresponds to  $|V_2|\sim 10^{-2}M_P^4$; \,
3)  q=0.1, which corresponds to  $|V_2|=10^{-4}M_P^4\approx (2.44\cdot 10^{17}GeV)^4$.
  For these three values of $|V_2|$, the plots of the effective TMT potential (\ref{Veff varphi tanh}) are shown in Fig.1. As can be seen, the larger $|V_2|$, the greater the plateau length, and this dependence is very sensitive to changes in $|V_2|$.

Note that if instead of the  scalar field model with the primordial potential (\ref{V m2phi2}) we choose a free massive scalar field $\phi$, i.e. $\lambda=0$, then, as can be seen from Eq.(\ref{Veff varphi tanh}), the  TMT  effective potential would have an infinite plateau.

For $\sqrt{\frac{2}{3}}\frac{\varphi}{M_P}>1$ we can restrict ourselves to the first degree of expansion in $e^{-\sqrt{\frac{2}{3}}\frac{\varphi}{M_P}}$ and find, using the chosen estimates of the model parameters
\begin{equation}
V_{eff}^{(0)}(\varphi) \approx
\frac{\lambda M_P^4}{4\xi^2(1+\zeta_v)}\left[1-8\left(1+\mathcal{O}\left(\frac{10^{-10}}{q^4}\right) \right)e^{-\sqrt{\frac{2}{3}}\frac{\varphi}{M_P}}
+\mathcal{O}\left(e^{-2\sqrt{\frac{2}{3}}\frac{\varphi}{M_P}}\right)\right].
\label{V in plateau}
\end{equation}
This is a representation of the potential, which, with a negligible correction (and for $\xi=\frac{1}{6}$), coincides with the corresponding expansion in $e^{-\sqrt{\frac{2}{3}}\frac{\varphi}{M_P}}$ of the potential $V(\varphi)\propto \tanh^{2n}(\frac{1}{\sqrt{6}}\frac{\varphi}{M_P})$ for n=2 found   in the  paper \cite{KL1} where it was named "the T-model". Thus, we can state that in the model we are studying, all inflationary predictions    coincide with the corresponding predictions of the T-model and, therefore, are consistent with Planck's data to the same extent. In particular, if $N$ is a number of remaining $e$-foldings of inflation at the time when the inflaton field $\varphi=\varphi(N)$, then $\varphi(N)$ is given by the equation (see e.g. \cite{KL1})
\begin{equation}
e^{2\sqrt{\xi}\frac{\varphi(N)}{M_P}}\approx 32\xi N.
\label{varphi N}
\end{equation}
For all parameters used in graphs of Fig.1, it follows that $\varphi(60)\approx 7.1M_P$.  

So, we found that when studying the simplest  field theory model (\ref{V m2phi2}) with nonminimal coupling to gravity in the framework of the TMT in the Palatini formalism, the TMT effective potential of the canonically normalized inflaton arises, which 1) has a plateau, the length of which is controlled by the parameter $V_2$; 2) 
when $\varphi$ exceeds a certain value depending on $V_2$, the potential passes from a flat shape to an almost exponential growth.
At first glance, it seems that the modification of the T-model potential obtained in the simplest model (\ref{S-gr-infl}) is reduced only to the appearance of an exponentially growing "tail". In fact, this is only partly true. Here TMT presents us with another surprise, which will be discovered and explored in the next subsections.

\subsection{The key role of the condition $\Upsilon(x)\neq 0$ as  a TMT unique attribute }

A solution (\ref{var varphi}) of Eq.(\ref{varphiB}) exists under the condition $\Upsilon(x)\neq 0$, Eq.(\ref{Phi neq 0}), i.e, 
only if $\Upsilon(x)>0$ or only if $\Upsilon(x)<0$. Therefore,  only those solutions of the system of equations obtained 
in the previous subsections  are valid for which {\em the corresponding $\Upsilon(x)$ is sign-definite}, since  Eq.(\ref{varphiB}) 
 is one of the equations of the system
\footnote
{Moreover, the solutions (\ref{var varphi}) for $\Upsilon(x)>0$ and for $\Upsilon(x)< 0$ generally contain different integration constants; the case of solutions with equal  integration constants  is extremely improbable. 
In this regard, it is worth noting here the difference between the model under study, which uses the volume form $dV_{\Upsilon}=\Upsilon d^4x$, Eq.(\ref{Phi}), defined in terms of  four scalar functions $\varphi_a$, from models that use the generally covariant measure of integration $dV_{\Phi}=\Phi d^4x $ with volume measure density of the form
\begin{equation}
\Phi=\frac{1}{3!}\epsilon^{\mu\nu\alpha\beta}\partial_{\mu}A_{\nu\alpha\beta},
\nonumber
\end{equation} 
where $A_{\nu\alpha\beta}$ is an auxiliary 3-index antisymmetric tensor gauge field. If the original action contains the term  $\int L\Phi d^4x$ with the corresponding Lagrangian $L$, then the variation with respect to  $A_{\nu\alpha\beta}$ gives the solution   $L=\mathcal{M}=const.$ without
restrictions on the sign of $\Phi$. Such an approach to choosing an alternative non-Riemannian measure of integration was used, for example, in models of Refs.\cite{BG1}-\cite{BG3}, where the original actions contain more than one term of this kind with different 3-index antisymmetric tensor gauge fields.
In such models, the scalar $\frac{\Phi}{\sqrt{-g}}$ also appears, which plays a role similar to that of the scalar $\zeta=\frac{\Upsilon}{\sqrt{-g }}$ in the model studied in this paper. But the essential difference of the model with the volume element $dV_{\Phi}=\Phi d^4x$ is that there is no need to require that the density $\Phi$ of the volume measure $dV_{\Phi}$ be nonvanishing. The result of this is that a single integration constant $\mathcal{M}$ can be chosen for a global solution applicable in the entire space-time manifold. This type of model can be used to construct a non-singular emergent universe followed by inflation, as in Ref. \cite{GHer 2023}, which continues the many  attempts  \cite{Emergent1}-\cite{Emergent4},\cite{Emergent} to implement the idea of a singularity-free inflationary universe. For further discussion of the possibility of a non-singular emergent universe, see, for example,  Refs. \cite{Wil vs.} and \cite{Past-completeness}. 
}
. It should be noted here that, as usual, by default we assume that the original metric $g_{\mu\nu}$ in the primordial action is regular, that is  $g=\det(g_{\mu\nu})<0$. 
Therefore, the validity of the solution regarding the fulfillment of the condition on the sign of $\Upsilon$ can be controlled by checking the sign of the scalar $\zeta=\Upsilon/\sqrt{-g}$.
In particular,  in the model under consideration, with our choice of parameters and constant of integration, the value of $\zeta$ in vacuum, Eq(\ref{zeta in vac 0}), is positive:  $\zeta_v>0$. Thus, {\bf only those cosmological solutions (together with their initial conditions)  are valid for which 
$\zeta(x)$ is positive throughout the evolution of the universe.} However, these solutions  lose their validity when we try to extend them to the values of $\varphi$ where $\zeta$ crosses zero and becomes negative. 

\begin{figure}
\includegraphics[width=13.0cm,height=8cm]{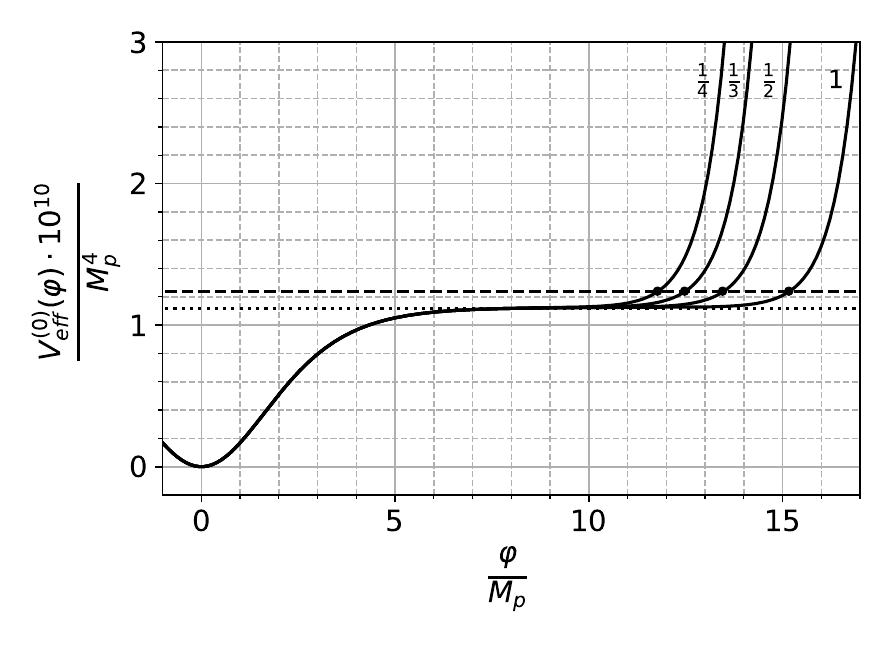}
\caption{
Four graphs of  $V_{eff}^{(0)}(\varphi)$, each curve is labeled by the corresponding value of $q=1$, $q=\frac{1}{2}$, $q=\frac{1}{3}$, $q=\frac{1}{4}$; 
the other parameters are the same for all curves: $\xi=\frac{1}{6}$, \, $\zeta_v=0.2$, \, $\lambda= 1.5\cdot 10^{-11}$, \, $m\approx 1.6\cdot 10^{13}GeV$ . The points of intersection of the dashed line with the function curves correspond to $\varphi=\varphi_*$ defined by 
Eq.(\ref{sinh 4 max fine tuned case}) and the values of $V_{eff}^{(0)}(\varphi_*)$ do not depend on $q$. But now $V_{eff}^{(0)}(\varphi_*)$ exceeds the plateau height by approximately 1.1 times instead of 1.5 in Fig.1. As a result, inflation can be driven from the very beginning by a scalar field with a potential, the shape of which differs little from a flat one.
This is achieved by choosing $\zeta_v=0.2$, which corresponds to choosing $|V_1|\approx 0.04|V_2|\approx (0.45qM_P)^4$. }
\label{fig2}
\end{figure}

If we assume that $\zeta(\varphi)$ vanishes at some value of $\varphi=\varphi_*$, then this situation is of special interest, and a significant part of what follows will be devoted to its study.
First of all, we should note that a more precise mathematical formulation is required. 
To understand the essence of the problem, we can restrict ourselves to the simplest model that we are currently studying, where $\zeta$ determined by the constraint  (\ref{zeta fine tun}) depends only on $\varphi$.
But the reasoning given below is also valid in more general models, e.g. studied in Sec.3 and discussed in Sec.4.  Let us assume that the solution of the field equations contains such a value of the field $\varphi=\varphi_*$ for which a formal substitution into the constraint (\ref{zeta fine tun}) gives $\zeta=0$. But strictly speaking, we have no right to do this substitution  because  the constraint (\ref{zeta fine tun}) was obtained under the condition $\Upsilon\neq 0$
which is equivalent to $\zeta\neq 0$. Consequetly, strictly speaking,  we can only describe the situation using the right-hand limit $\zeta\rightarrow 0^{\,+}$ (and, of course,  $\Upsilon(x)\rightarrow 0^{\,+}$), i.e.
\begin{equation}
\lim_{\varphi \to \varphi_*^{\,-}}\zeta=0.
\label{lim zeta star}
\end{equation}
No less interesting is the question of the left-hand limit giving $\zeta\rightarrow 0^{\,-}$ (and, of course,  $\Upsilon(x)\rightarrow 0^{\,-}$). The answer and its consequences turn out to be very non-trivial and will be discussed in Sec.4.
But until then, we will focus on possible initial conditions for inflationary solutions to the cosmological equations. {\it It follows from the above that those initial conditions for which $\zeta\leq 0$ should be excluded from consideration as an artifact in our Universe.}
As we will see in the next subsection, this is exactly what happens in the model under study.

\subsection{Strict bound on maximum initial value of $\varphi$ 
\\
imposed by the condition $\zeta\equiv\frac{\Upsilon}{\sqrt{-g}} >0$ in the simplest model}

Let us start with Ref.\cite{singular alpha}, where A. Linde showed that in the singular $\alpha$-attractor model one can obtain a potential similar to the one in Fig.1.
If $\alpha >1/3$ and the initial value of 
$\varphi$ is sufficiently large, then due to the almost 
exponential form of the potential, the expansion of the universe can begin with a power-law inflation.
Therefore, according to Linde's idea, inflation "may begin already at the Planck density, which solves the problem of initial conditions in this class of models along the lines of \cite{Linde 1985}".

It turns out that this idea cannot be implemented in the TMT models we are studying.  To understand the reason, we must return to the conclusion made in the previous subsection based on the analysis of the unique capabilities of TMT.
Using the constraint (\ref{zeta fine tun}), one can see that the situation described by the limit (\ref{lim zeta star})
is actually realized and
 $\varphi_*$ is  determined by the algebraic equation
\begin{equation}
\frac{1}{\xi}\frac{m^2}{M_P^2}{\sinh}^2\sqrt{\xi}\frac{\varphi_*}{M_P}+\frac{\lambda}{2\xi^2}\sinh^4\sqrt{\xi}\frac{\varphi_*}{M_P}=
4q^4\zeta_v(1+\zeta_v)
\label{condition zeta 0 fine tuned case}
\end{equation}
With our parameter estimates $\frac{m^2}{M_P^2}$ and $\frac{\lambda}{2\xi}$ are of the same order. Since we are interested in the region 
$\sqrt{\xi}\frac{\varphi_*}{M_P}\gg 1$, the first term on the left  side of Eq.(\ref{condition zeta 0 fine tuned case}) is negligible compared to the second, and we find for $\varphi_*$
\begin{equation}
\sinh^4\sqrt{\xi}\frac{\varphi_*}{M_P}\approx\frac{1}{16}e^{4\sqrt{\xi}\frac{\varphi_*}{M_P}}=\frac{8q^4\xi^2\zeta_v(1+\zeta_v)}{\lambda}
\label{sinh 4 max fine tuned case}
\end{equation}
In Fig.1, the points on the plots of $V_{eff}^{(0)}$ corresponding to $\varphi_*$ are marked with dots.
 Hence $\zeta >0$ for any $\varphi<\varphi_*$,  i.e. 
$\varphi_*$ is the limiting upper bound of admissible  values of $\varphi$.  Thus,  for each of the curves in Fig. 1,  {\bf the part of the TMT effective potential corresponding to the interval $\varphi\geq\varphi_*$  is an artifact in our Universe}. The TMT effective potential $V_{eff}^{(0)}(\varphi)$ remains finite as $\varphi\rightarrow \varphi_*^{\,-}$, and it follows from the structure of the TMT effective energy-momentum tensor that, in any proximity to the hypersurface
 $\zeta(x)=0$ it has no singularities.

Inserting (\ref{sinh 4 max fine tuned case}) into the effective potential (\ref{Veff varphi tanh}) we obtain
\begin{equation}
V_{eff}^{(0)}(\varphi)|_{\zeta\approx 0} \approx \lim_{\varphi \to \varphi_*^{\,-}}V_{eff}^{(0)}(\varphi)\approx
\frac{\lambda M_P^4}{4\xi^2(1+\zeta_v)}\left(1+\frac{1}{2}\zeta_v\right)=\left(1+\frac{1}{2}\zeta_v\right)V_{eff}^{(0)}(\varphi)|_{plateau},
\label{Veff as z = 0}
\end{equation}
where $V_{eff}^{(0)}(\varphi)|_{plateau}$ is the typical hight of the plateau of the TMT effective potential at 
$1\ll \sqrt{\xi}\frac{\varphi}{M_P}<\sqrt{\xi}\frac{\varphi_*}{M_P}$. With the chosen parameters $\zeta_v=\sqrt{V_1/V_2}\leq 1$, and we conclude that for the initial value 
$\varphi_{in}$ arbitrarily close to the limiting value $\varphi_*$ (admissible by the condition $\zeta > 0$), the ratio of $V_{eff }^ {( 0 )}(\varphi_{in})$ to the typical height of the plateau does not exceed 1.5 and the value of this ratio does not depend on $|V_2|=(qM_P)^4$.
 As can be seen from the graphs 
in Fig.1, where  $\zeta_v=1$ is  chosen, the evolution can begin when the effective TMT potential $V_{eff }^{( 0)}(\varphi)$  has an almost exponentially growing shape. 
But in the TMT model under study, $\varphi_{in}$ cannot exceed $\varphi_*$, and, consequently, the TMT effective  potential cannot exceed the value
$V_{eff }^{( 0)}(\varphi_*)\ll M_P^4$. Therefore, the scenario proposed by Linde for solving the problem of the initial conditions  is not realizable in the model under study.

As we will see, the more general model studied in Sec.3 may provide an alternative approach to solving the problem of initial conditions for inflation.
However, here we restrict ourselves to demonstrating a simple way to avoid complicating the inflationary scenario caused by the almost exponentially growing part of  $V_{eff }^{(0)}(\varphi)$. To do this, it is enough to use arbitrariness in the choice of the parameter $|V_1|$ or, what is the same, in the choice of $\zeta_v$, see Eq.(\ref{zeta in vac 0}). For example, if $\zeta_v = 0.2$, then $V_{eff }^{(0)}(\varphi_*)$ exceeds the typical plateau height by about $10\%$. 
In this case, when 
$\varphi_{in}$ is close to $\varphi_*$, inflation is driven from the very beginning by a scalar field with a potential, the shape of which differs little from a flat one.
As an example with $\zeta_v = 0.2$, the  plots of the TMT effective potentials $V_{eff }^{(0)}(\varphi)$ for four values of $q$  are shown in Fig.2, where the points on the graphs  corresponding to $q$-dependent values of $\varphi_*$  are marked. Note that the value $\zeta_v = 0.2$ corresponds to the choice of $|V_1|\approx 0.04|V_2|\approx (0.45qM_P)^4$.

The qualitative conclusion following from these results is as follows: {\em within the framework of the considered simplest model, there is a wide range of model parameters, where the theory forbids in our Universe the canonically normalized inflaton field $\varphi$ to exceed the value, starting from which the potential has an exponentially growing "tail"}.
 Note that, as will be shown in Sec.3 and Appendix B, this conclusion can be violated if the initial gradient energy density is greater than the initial kinetic energy density.

To complete the picture, we still need to make sure that  $\zeta$ remains positive  during the entire time of cosmological evolution.  To control the change in $\zeta$ we need to know the sign of $\frac{d\zeta}{d\varphi}$.
  It follows from Eq.(\ref{phi via canonical varphi}) that: 1) as $\varphi >0$ also $\phi >0$; 2) $\frac{d\phi}{d\varphi}>0$;
 3)  $sign\left(\frac{d\zeta}{d\varphi}\right)=sign\left(\frac{d\zeta}{d\phi}\right)$. Therefore, to simplify  the calculation,
 instead of  $\frac{d\zeta}{d\varphi}$  one can find 
$\frac{d\zeta}{d\phi}$. 
Using Eqs.(\ref{V1 V2 M}) and (\ref{zeta in vac 0}) we obtain from the constraint (\ref{Constraint})
\begin{equation}
\frac{d\zeta}{d\phi}=-\frac{2|V_2|(1+\zeta_v)^2}
{\left[2|V_2|(1+\zeta_v)+V(\phi)\right]^2}V^{\prime}(\phi).
\label{dz dphi in lambda 0}
\end{equation}
Eq.(\ref{V m2phi2}) shows that $V^{\prime}(\phi)>0$ for $\phi >0$ and hence $\frac{d\zeta}{d\phi}<0$. 
Therefore, in the inflation process governed by a monotonically decreasing classical scalar field $\varphi >0$, $\zeta(\varphi)$ increases monotonically. The initial value $\varphi_{in}$, from which inflation starts, can be arbitrarily close to $\varphi_*$, but it must be $\varphi_{in}<\varphi_*$.
Accordingly, $\zeta(\varphi(t))$ can start with a positive initial value $\zeta(\varphi_{in})>0$, arbitrarily close to zero
\begin{equation}
\zeta(\varphi_{in})\rightarrow 0^+ \quad \text{as} \quad  \varphi_{in}\rightarrow\varphi_*^{\,\,\, -},
\label{zin 0}
\end{equation}
and $\zeta(\varphi)$ increases monotonically during inflation. 
After the end of inflation, in the process of transition to the vacuum state, oscillations of $\varphi$ can cause oscillations of $\zeta$, but it remains positive and approaches its vacuum value $\zeta_v>0$.

\section{A more general model and natural TMT constraints on
\\
 the   initial kinetic and gradient energy densities}

A naive attempt to combine the constraint on the height of the potential of plateau-like models  imposed by  data of 
recent cosmological observations
 \cite{Planck}-\cite{BICEP and Keck} with the constraints on the initial kinetic and gradient energy densities (\ref{Cond for begin infl}) necessary for the onset of inflation leads us to the need to ensure that the following inequalities hold 
\begin{eqnarray}
\rho_{kin,in}&=&\frac{1}{2}\dot{\varphi}_{in}^2\lesssim V_{eff}^{(0)}=\mathcal{O}(1)\cdot 10^{-10}M_P^4,
\nonumber
\\
\rho_{grad,in}&=&\frac{1}{2}|(\partial^k\varphi)_{in}(\partial_{k}\varphi)_{in}|\lesssim V_{eff}^{(0)}=\mathcal{O}(1)\cdot  10^{-10}M_P^4.
\label{kin and grad less V}
\end{eqnarray}
But this is precisely what was the main object of justified criticism in Ref.\cite{Stein} that we mentioned in Introduction.
However, in the model formulated in this section, which is somewhat more general than the model in Sec.2, we will show that there is an interval of initial values $\varphi_{in}$ where, along with the condition $\zeta> 0$, it is also guaranteed  fulfillment of the conditions (\ref{kin and grad less V}) necessary for the onset of inflation.

\subsection{A more general model}

As already mentioned in Sec.2.1, in the general case $\Upsilon$ and $\sqrt{-g}$ can enter the volume element with arbitrary coefficients. The choice of $(\sqrt{-g}+\Upsilon)d^4x$ as the volume element in all gravity and matter  terms, made in Sec.2.1, means that all results were obtained there with the most simplified approach to the choice of these coefficients. Therefore, it would be interesting to know what other new results can be obtained by abandoning such a simplification of the model. Bearing in mind the relevant discussion in Sec.2.1, we choose volume elements in the form $\left(b_i\sqrt{-g}+\Upsilon\right)d^4x$, where $b_i$ is the model parameter corresponding to the i's term in the primordial action.

Generalizing in this way the model studied so far, the primordial action can be represented as follows
\begin{eqnarray}
S&=&\int d^4x\left[-\frac{M_P^2}{2}(\sqrt{-g}+\Upsilon)\left(1+\xi \frac{\phi^2}{M_P^2}\right)R(\Gamma,g)\right]
\nonumber
\\
&+&\int d^4x\left[(b_k\sqrt{-g}+\Upsilon) \frac{1}{2}g^{\alpha\beta}\phi_{,\alpha}\phi_{,\beta} 
- (b_p\sqrt{-g}+\Upsilon)\left(\frac{1}{2}m^2\phi^2+\frac{\lambda}{4}\phi^4\right) \right]
\nonumber
\\
&-&\int d^4x\left[\sqrt{-g}V_1+\frac{\Upsilon^2}{\sqrt{-g}}V_2\right],
\label{S without fine tun}
\end{eqnarray}
where $b_k$ and $b_p$ are additional model parameters. Of course, this is not the most general action of this kind, because another parameter of the same type can be added to the volume element, coming with a non-minimal coupling. One can believe that the reason for the deviation of the parameters $b_k$ and $b_p$ from unity  is quantum corrections.
 Then it is natural to assume that these corrections are small, and for further consideration we choose $0<b_k<1$ and $0.5<b_p<1$. But it is worth noting that in the calculations, the results of which are given below, the possible smallness of $1-b_k$ and $1-b_p$ is not used.

It is known \cite{GK3} that in TMT models with $b_k\neq 1$ and $b_p\neq 1$, the implementation of all seven steps of the TMT procedure listed in Appendix A leads to a TMT effective  action with a structure typical for K-essence models \cite{K-essence1}-\cite{K-essence5}.
Therefore, repeating again the TMT procedure in model (\ref{S without fine tun}), we quite expectedly come to a similar result. 

Following the prescription of the TMT procedure and similarly to what was done in Sec.2, we have to consider the equations of motion following from the primordial action (\ref{S without fine tun}).
Varying the action with respect to
   scalar functions $\varphi_{a}$, from which $\Upsilon$ is  built, we obtain an equation that coincides with Eq.(\ref{varphiB}). Its solution is the same as in  Eq.(\ref{var varphi})  if the same condition (\ref{Phi neq 0}) is satisfied, i.e. everywhere $\Upsilon(x)\neq 0$. But varying the action (\ref{S without fine tun}) with respect to the metric tensor $g_{\mu\nu}$ leads to equations that differ from Eq.(\ref{Grav.eq})  by the presence of additional parameters $b_k$ and $b_p $. The requirement of the consistency of the obtained equations,  unlike Eq.(\ref{Constraint}), now  defines the scalar $\zeta(x)$ as a function which depends  not only on $\phi$, but also on  $g^{\alpha\beta}\phi_{,\alpha}\phi_{,\beta}$. That is why  it is more convenient firstly to keep it in the form
\begin{equation}
\zeta\left[{\mathcal M}-2V_2+V(\phi)+(1-b_k)\frac{1}{2}g^{\alpha\beta}\phi_{,\alpha}\phi_{,\beta}\right]
-{\mathcal M}+2V_1+(2b_p-1)V(\phi)+(1-b_k)\frac{1}{2}g^{\alpha\beta}\phi_{,\alpha}\phi_{,\beta}=0
\label{zeta of phi and X firstly}
\end{equation}
and to represent in the final form  only after transition to the Einstein frame. The latter is performed as in Eq.(\ref{gmunuEin}) where now 
$\zeta=\zeta(\phi, g^{\alpha\beta}\phi_{,\alpha}\phi_{,\beta})$. Thus,  the constraint in the Einstein frame  in the studying now model  reads as follows 
\begin{equation}
\zeta(\phi,X_{\phi})=\frac{{\mathcal M}+2|V_1|-(2b_p-1)V(\phi)
-(1-b_k)\left(1+\xi\frac{\phi^2}{M_P^2}\right)\cdot X_{\phi}}
{{\mathcal M}+2|V_2|+V(\phi)+(1-b_k)\left(1+\xi\frac{\phi^2}{M_P^2}\right)\cdot X_{\phi}},  \qquad
 X_{\phi}=\frac{1}{2}{\tilde g}^{\alpha\beta}\phi_{,\alpha}\phi_{,\beta}
\label{1+ zeta no fine tun phi M}
\end{equation}
The $\phi$-equation also contains parameters $b_k$ and $b_p$. Instead of  the function $U_{eff}(\phi,\zeta(\phi);{\mathcal M})$ obtained in  Sec.2 and defined by Eq.(\ref{u zeta}), in the model with the action (\ref{S without fine tun})  we obtain 
\begin{equation}
U_{eff}(\phi,\zeta(\phi,X_{\phi});{\mathcal M})=
\frac{1}{\left(1+\xi\frac{\phi^2}{M_P^2}\right)^2}\left[\frac{{\mathcal M}+|V_1|+|V_2|+(1-b_p)V(\phi)}{[1+\zeta(\phi,X_{\phi})]^2}-|V_2|\right],
 \label{u zeta model sec 4}
\end{equation}

In the space-time region, where $\Upsilon>0$, we again choose the integration constant ${\mathcal M}$ as in Sec.2.2, i.e. according to Eq.(\ref{V1 V2 M}), which provides zero vacuum energy density.
 The last fact will be marked with $(0)$ in the notation of all relevant quantities.  After passing  to the Einstein frame  in all other equations and using the redifinition (\ref{phi via canonical varphi}) to the canonically normalized scalar field $\varphi$,  the expression for the TMT effective action takes the form
\begin{equation}
S_{eff}^{(0)}=\int \left[-\frac{M_P^2}{2}R(\tilde{g})+
L_{eff}^{(0)}\left(\phi(\varphi),\zeta(\varphi,X_{\varphi})\right)\right]
\sqrt{-\tilde{g}}d^4x, 
\label{eff action general via zeta}
\end{equation}
where the TMT effective Lagrangian for the scalar field $\varphi$ appears as following
\begin{equation}
L_{eff}^{(0)}(\phi(\varphi),\zeta(\varphi,X_{\varphi}))=X_{\varphi}-U_{eff}^{(0)}(\phi(\varphi),\zeta(\varphi,X_{\varphi})),
 \qquad X_{\varphi}=\frac{1}{2}\tilde{g}^{\alpha\beta}\varphi_{,\alpha}\varphi_{,\beta}
\label{eff L general via zeta}
\end{equation}
and $U_{eff}^{(0)}(\phi(\varphi),\zeta(\varphi,X_{\varphi}))$  in terms of $\zeta=\zeta(\varphi,X_{\varphi})$ reads
\begin{equation}
U_{eff}^{(0)}(\phi(\varphi),\zeta(\varphi,X_{\varphi}))=\frac{1}{\cosh^4z}\left[\frac{|V_2|(1+\zeta_v)^2+(1-b_p)V(\phi(\varphi))}{(1+\zeta)^2}-|V_2|\right],
\label{u eff  via zeta}
\end{equation}
where $z=\sqrt{\xi}\frac{\varphi}{M_P}$.
Here $\zeta=\zeta(\varphi,X_{\varphi})$ is determined by the constraint following from Eq.(\ref{1+ zeta no fine tun phi M}) in which we have chosen the integration constant ${\mathcal M}_0=2\sqrt{V_1V_2}$ and used a redefinition  (\ref{phi via canonical varphi}) to a canonically normalized scalar field 
$\varphi$:
\begin{equation}
\zeta(\varphi,X_{\varphi})=\frac{2|V_2|\zeta_v(1+\zeta_v)-(2b_p-1)\left[\frac{m^2M_P^2}{2\xi}\sinh^2z+\frac{\lambda}{4\xi^2}M_P^4\sinh^4z\right]
-(1-b_k)\cosh^4z\cdot X_{\varphi}}
{2|V_2|(1+\zeta_v)+\frac{m^2M_P^2}{2\xi}\sinh^2z+\frac{\lambda}{4\xi^2}M_P^4\sinh^4z+(1-b_k)\cosh^4z\cdot X_{\varphi}}
\label{1+ zeta no fine tun}
\end{equation}
A significant difference from the simplest model studied in Sec.2 is that now $\zeta$ turns out to be a function depending not only on $\varphi$ but also 
on $X_{\varphi}$.

Inserting $\zeta(\varphi,X_{\varphi})$ to Eq.(\ref{u eff  via zeta}) and using Eqs.(\ref{V m2phi2}) and (\ref{phi via canonical varphi})
 we obtain the final expression for the TMT effective  Lagrangian
of scalar field $\varphi$,
 $L_{eff}^{(0)}(\varphi,X_{\varphi})\equiv L_{eff}^{(0)}(\phi(\varphi),\zeta(\varphi,X_{\varphi}))$, which can be represented in the following  form, typical for K-essence models \cite{K-essence1}-\cite{K-essence5}
\begin{equation}
L_{eff}^{(0)}(\varphi,X_{\varphi})=X_{\varphi}-V_{eff}^{(0)}(\varphi)-K_1(\varphi)X_{\varphi}-K_2(\varphi)\frac{X_{\varphi}^2}{M_P^4},
\label{Leff final no fine tun}
\end{equation}
where
\begin{equation}
V_{eff}^{(0)}(\varphi) =\frac{M_P^4}{4\xi^2}{\tanh}^4z\cdot F(z),
\label{Veff varphi tanh no fine tun 1}
\end{equation}
\begin{equation}
F(z)=\frac{\lambda q^4(\zeta_v+b_p)+
\frac{m^4}{4M_P^4}+
\frac{\lambda}{4\xi}\frac{m^2}{M_P^2}{\sinh}^2z+\frac{\lambda^2}{16\xi^2}\sinh^4z+
2\xi q^4(\zeta_v+b_p)\frac{m^2}{M_P^2}\cdot \sinh^{-2}z }
{(1+\zeta_v)^2q^4+(1-b_p)\left(\frac{1}{2\xi}\frac{m^2}{M_P^2}\sinh^2z+\frac{\lambda}{4\xi^2}{\sinh}^4z\right)},
\label{Veff varphi tanh no fine tun 2}
\end{equation}

\begin{equation}
K_1(\varphi)=\frac{1-b_k}{2\cosh^2z}\, \cdot\frac{8q^4\xi^2(1+\zeta_v)+2\xi\frac{m^2}{M_P^2}\cdot \sinh^{2}z+\lambda\sinh^4z}
{4q^4\xi^2(1+\zeta_v)^2+2(1-b_p)\xi\frac{m^2}{M_P^2}\cdot \sinh^{2}z+\lambda(1-b_p)\sinh^4z},
\label{K1}
\end{equation}
\begin{equation}
K_2(\varphi)=\frac{(1-b_k)^2}{4\left[q^4(1+\zeta_v)^2+\frac{1-b_p}{2\xi}\frac{m^2}{M_P^2}\cdot \sinh^{2}z
+\frac{\lambda(1-b_p)}{4\xi^2}\sinh^4z\right]}.
\label{K2}
\end{equation}

As can be seen from Eqs.(\ref{Veff varphi tanh no fine tun 1}), (\ref{Veff varphi tanh no fine tun 2}), the choice of $b_p\neq 1$ radically changes the behavior of $V_{eff}^{(0)}(\varphi)$ at $\varphi \ggg M_P$.  Fig.3 and Fig.4 show two examples where instead of almost exponential unlimited growth, which was at $b_p=1$, a second plateau appears. In Fig.3, where $b_p=0.7$ is chosen, the second plateau is about 1.33 times higher than the first plateau. But on the graph of $V_{eff}^{(0)}(\varphi)$ in Fig.4, due to the choice of $b_p\approx 0.5$, the height of the second plateau exceeds the height of the first plateau by about 1.03 times.

If $X_{\varphi }= 0$, then, as in the case of $b_p=1$ in Sec.2, there  exists a value  $\varphi_0$ such that the infinite interval  $\varphi>\varphi_0$ is also an artifact.
To show this, we first note, using the constraint (\ref{1+ zeta no fine tun}), that at the minimum of $V_{eff}^{(0)}(\varphi)$, i.e. for $ \varphi(x)\equiv 0$, scalar $\zeta$ has the same value $\zeta_v >0$ as defined by Eq.(\ref{zeta in vac 0}). Therefore, throughout the entire process of cosmological evolution, $\zeta$ must be positive. This was of paramount importance in the previous section and will be equally important for all subsequent analysis. Now, given the constraint (\ref{1+ zeta no fine tun}) when $X_{\varphi }= 0$, we see that   $\zeta\rightarrow 0^+$ when $\varphi\rightarrow \varphi_0^{\,\,\,-}$ 
where $\varphi_0$ is defined by the relation 
\begin{equation}
\sinh^4\sqrt{\xi}\frac{\varphi_0}{M_P}=\frac{8q^4\xi^2\zeta_v(1+\zeta_v)}{\lambda (2b_p-1)}
\label{sinh 4 max no fine tuned case}
\end{equation}
obtained after neglecting a very small correction from the  term $\propto \sinh^2\sqrt{\xi}\frac{\varphi_0}{M_P}$.
This is a generalization of the definition of  $\varphi_*$, Eq.(\ref{sinh 4 max fine tuned case}), to the case  $b_p\neq1$. For the parameters used in the plot in   Fig.3, relation (\ref{sinh 4 max no fine tuned case}) gives $\varphi_0\approx 12.9M_P$. For the parameters used in the plot in   Fig.4, relation (\ref{sinh 4 max no fine tuned case}) gives $\varphi_0\approx 13.7M_P$. The corresponding points on the curves are marked with dots.  

\begin{figure}
\includegraphics[width=13.0cm,height=8cm]{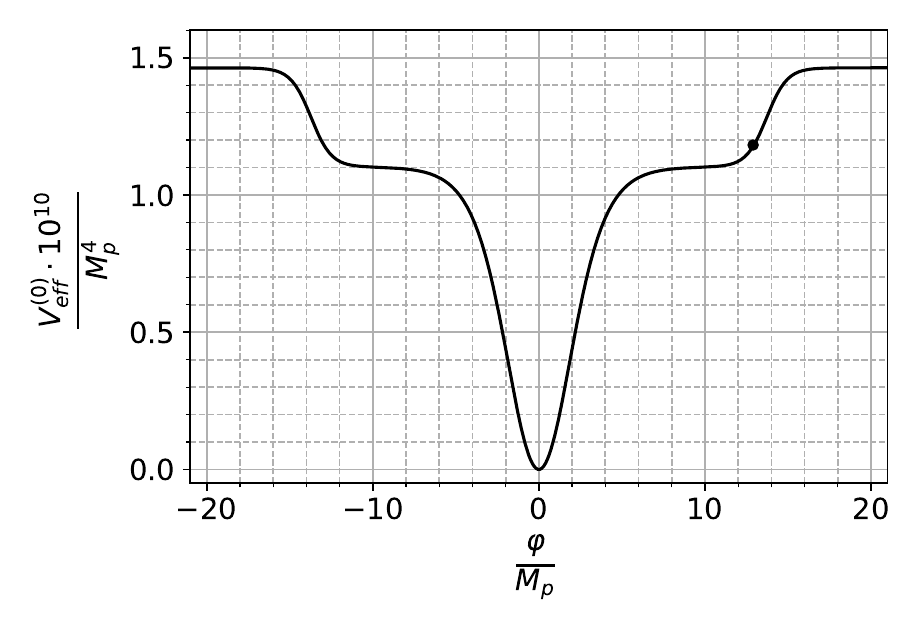}
\caption{In the model with $b_p\neq 0$, $V_{eff}^{(0)}(\varphi)$ can, generally speaking, have two plateaus. Here, the following parameters have been chosen as an example:
  $q=\frac{1}{3}$, \, $\xi=\frac{1}{6}$, \, $\zeta_v=0.2$, \, $b_p=b_k=0.7$, \, $\lambda\approx 2\cdot 10^{-11}$, \,  $m\approx 2\cdot 10^{13}GeV$. The dot on the plot corresponds to $\varphi_0\approx 12.9 M_P$ defined by relation (\ref{sinh 4 max no fine tuned case}).
In this case $\varphi_{in}^{(min)}$ given by Eq.(\ref{interval phi for X positive}) is  $\varphi_{in}^{(min)}\approx 12.6 M_P$ (corresponding point is not marked here).}
\label{fig3}
\end{figure}

\begin{figure}
\includegraphics[width=13.0cm,height=8cm]{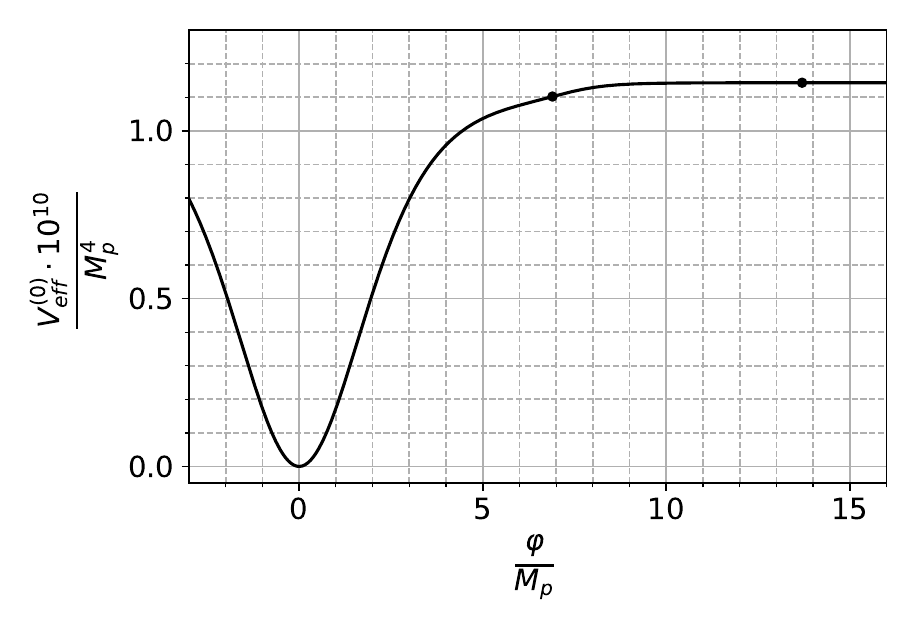}
\caption{Plot of $V_{eff}^{(0)}(\varphi)$ for the case of the following choice of parameters:
  $q^4=10^{-6}$, \, $\xi=0.18=\frac{1}{6\alpha}$, that is $\alpha=0.926$, \, $\zeta_v=0.2$, \, $b_p=0.5+10^{-6}$, \, $b_k=0.5$, \, $\lambda= 2.9\cdot 10^{-11}$, \,  $m\approx 2\cdot 10^{13}GeV$. The two dots on the plot correspond to $\varphi_0\approx 13.7 M_P$ 
defined by relation (\ref{sinh 4 max no fine tuned case}) and
$\varphi_{in}^{(min)}\approx 6.8 M_P$ defined by Eq.(\ref{interval phi for X positive}).}
\label{fig4}
\end{figure}

\subsection{TMT constraints on the initial kinetic and gradient energy densities
\\
 and initial conditions for inflation}

The standard, non-TMT, formulation of the initial conditions for inflation driven by a scalar field includes specifying or at least estimating the initial value of the field $\varphi_{in}$ and its first derivatives or, equivalently, the initial kinetic energy density  $\rho_{kin,in}$ and the gradient energy density $\rho_{grad,in}$.
It is fundamentally important and taken for granted that there is no dependence in any form between  $\varphi_{in}$ and $\rho_{kin,in}$, as well as between   $\varphi_{in}$ and $\rho_{grad,in}$. In the simplest model of Sec.2, although there is a restriction on the initial values of $\varphi_{in}<\varphi_*$ (following from the condition $\zeta>0$), no restrictions on $\rho_{kin,in}$ and $\rho_{grad,in}$ appear.
 But in the more general model with action  (\ref{S without fine tun}), {\em the scalar $\zeta$ given by the constraint (\ref{1+ zeta no fine tun}) turns out to depend not only on $\varphi$, but also on $X_{\varphi}=\frac{1}{2}\tilde{g}^{\alpha\beta}\varphi_{,\alpha}\varphi_{,\beta}$}. As a result, the condition $\zeta>0$ imposes restrictions in the form of inequalities on the admissible ranges of $\varphi$ {\bf and} $X_{\varphi}$. We are interested in the constraints on the initial values $\varphi_{in}$ and $X_{\varphi}^{(in)}$ imposed {\em  both}  by condition $\zeta(\varphi_{in}, X_{\varphi }^{(in )})>0$ {\em and} upper bounds on $\rho_{kin,in}$ {\em and} $\rho_{grad,in}$, Eq.(\ref{kin and grad less V}).
 Inflation is possible only if all these conditions are met together.

 The results of the requirement that $\zeta(\varphi, X_{\varphi})$ be positive can be obtained from a detailed study of the inequality
\begin{equation}
\frac{2|V_2|\zeta_v(1+\zeta_v)-(2b_p-1)\left[\frac{m^2M_P^2}{2\xi}\sinh^2z+\frac{\lambda}{4\xi^2}M_P^4\sinh^4z\right]
-(1-b_k)\cosh^4z\cdot X_{\varphi}}
{2|V_2|(1+\zeta_v)+\frac{m^2M_P^2}{2\xi}\sinh^2z+\frac{\lambda}{4\xi^2}M_P^4\sinh^4z+(1-b_k)\cosh^4z\cdot X_{\varphi}}>0,
\label{1+ zeta no fine tun larger 0}
\end{equation}
which is carried out in Appendix B.
But in doing this analysis, we will have to take into account the significant difference between how kinetic $\rho_{kin}$ and gradient $\rho_{grad}$ energy densities enter Einstein's equations and the constraint (\ref{1+ zeta no fine tun}) (and hence the inequality (\ref{1+ zeta no fine tun larger 0})). Indeed, while 
the kinetic and gradient energy densities enter the Einstein equations in the form of a sum, they enter the constraint  in the form 
\begin{equation}
X_{\varphi}=\frac{1}{2}\tilde{g}^{\alpha\beta}\varphi_{,\alpha}\varphi_{,\beta}=\frac{1}{2}\left(\dot{\varphi}^2-
\frac{1}{a^2}(\nabla\varphi)^2\right)
=\rho_{kin}-\rho_{grad},
\label{X as difference}
\end{equation}
 that is, in fact, in the form of a difference. Therefore, 
$X_{\varphi}^{(in)}$ can be positive or negative depending on how inhomogeneous and anisotropic at the beginning of inflation was the space domain  whose expansion generates our Universe \footnote{The case when $\frac{1}{2}\dot{\varphi}_{in}^2> 10^{-10}M_P^4$ and $\frac{1}{2}|(\partial^k\varphi)_{in}(\partial_{k}\varphi)_{in}|>10^ {-10}M_P^4$, but $|X_{\varphi}^{(in)}|<10^{-10}M_P^4$ seems rather  improbable and we will not consider it.}. Since the sign of $X_{\varphi}^{(in)}$ can significantly affect the results obtained from the condition $\zeta>0$, the cases $X_{\varphi}^{(in)}>0$ and $X_{ \varphi}^{(in)}<0$ are examined separately in the Appendix B. The main results are summarized here.

\subsubsection
 {The main results in the case $X_{\varphi}^{(in)}>0$}

If $X_{\varphi}^{(in)}>0$, then there exists the minimal initial value  $\varphi_{in}^{(min)}$ defined by Eq.(\ref{interval phi for X positive}) such that
{\bf on the interval}
\begin{equation}
\varphi_{in}^{(min)}\leq  \varphi_{in}
<\varphi_0
\label{phi 0 and phi min}
\end{equation}
{\bf the condition $\zeta(\varphi_{in} , X_{\varphi}^{(in)})>0$  and the necessary conditions (\ref{kin and grad less V}) for the beginning of inflation are guaranteed to be fulfilled. Upper bound of admissible  values of  $X_{\varphi}^{(in)}$  tends to zero as $\varphi_{in}\rightarrow \varphi_0^{\,\, -}$. The condition $\zeta>0$ forbids $\varphi_{in}$ to be $\varphi_{in}\geq\varphi_0$.}

As examples,  consider two options presented in Figs. 3 and 4.

1)  For the parameters used in Fig.3,  we get  $12.6 M_P\approx\varphi_{in}^{(min)}\leq\varphi_{in}<\varphi_0\approx 12.9M_P$. In Sec.2.3, using Eq.(\ref{varphi N}) with $\xi=\frac{1}{6}$, we obtained  that $\varphi(N)$ correspoding to 60 $e$-foldings is $\varphi(60)\approx 7.1M_P$.
 One can check that  in the more general model of this section, Eq.(\ref{varphi N}) holds with the same approximation as in the simplest model of Sec.2.
 Therefore, in this case, $\varphi(60)\approx 7.1M_P<\varphi_{in}^{(min)}\approx12.6 M_P$, that is $\varphi(60)$ is outside the interval (\ref{phi 0 and phi min}).

2) For the set of parameters used in Fig.4, we get  $\varphi_{in}^{(min)}\approx 6.8 M_P\leq\varphi_{in}<\varphi_0\approx 13.7M_P$. In this case, a significant decrease in $\varphi_{in}^{(min)}$ and a corresponding numerical expansion of interval (\ref{phi 0 and phi min}) is achieved mainly due to the choice of $b_p=0.5+10^{-6}$. Note that the chosen value of $\xi =0.18$ can be represented as $\xi=\frac{1}{6\alpha}$ with $\alpha \approx 0.93$.  Now we get  that $\varphi(N)$ correspoding to 60 $e$-foldings is $\varphi(60)\approx 6.9M_P>\varphi_{in}^{(min)}\approx 6.8 M_P$, i.e. is inside the  interval (\ref{phi 0 and phi min}).

3) For $\varphi_{in}<\varphi_{in}^{(min)}$ the condition $\zeta(\varphi, X_{\varphi})>0$ is satisfied, but the conditions
 (\ref{kin and grad less V}) required for the onset of inflation  may or may not hold.

\subsubsection{Some additional important results of the model in the case $X_{\varphi}^{(in)}>0$}

\begin{itemize}

\item

{\em The effect of  the  terms $K_1(\varphi)X_{\varphi}+K_2(\varphi)\frac{X_{\varphi}^2}{M_P^4}$ in the TMT effective Lagrangian
in the case $X_{\varphi}^{(in)}>0$}

We must consider the effect of the last two terms in the TMT effective Lagrangian $L_{eff}^{(0)}(\varphi,X_{\varphi})$, Eq.(\ref{Leff final no fine tun}), in the interval described by 
Eq.(\ref{phi 0 and phi min}). 
With the parameters used in Fig.3, 
we estimate the maximum values of $K_1$ and $K_2$:  $K_{1, max}=K_1(\varphi_{in}^{(min)})\approx 4\cdot 10^{-5}$ and  $K_{2, max}=K_2(\varphi_{in}^{(min)})\approx  1.4$. So using also  Eqs.(\ref{restr on X from above final}) and (\ref{approx X to hight}), the contribution of the last two terms to the effective Lagrangian (\ref{Leff final no fine tun}) in the interval  
(\ref{phi 0 and phi min}) is estimated to be  
\begin{equation}
K_1(\varphi_{in})X_{\varphi}^{(in)}+K_2(\varphi_{in})\frac{(X_{\varphi}^{(in)})^2}{M_P^4}\lesssim 4\cdot 10^{-5}\cdot V_{eff}^{(0)}(\varphi_{in}).
\label{contr K1K2 main text}
\end{equation}
With the parameters used in Fig.4, 
similar estimates give:  $K_{1, max}=K_1(\varphi_{in}^{(min)})\approx 2\cdot 10^{-5}$ and  $K_{2, max}=K_2(\varphi_{in}^{(min)})\approx  2\cdot 10^{5}$. So using also  Eqs.(\ref{restr on X from above final}) and (\ref{approx X to hight}), the contribution of the last two terms to the effective Lagrangian (\ref{Leff final no fine tun})  in the interval  
(\ref{phi 0 and phi min}) is estimated to be  the same as  in Eq.(\ref{contr K1K2 main text}).
Thus in both of the above examples, 
 the contribution of the last two terms to the effective Lagrangian (\ref{Leff final no fine tun})  in the interval  
(\ref{phi 0 and phi min}) is negligible if we take into account the absence of any reliable data on the beginning of inflation.

\item

{\em
Estimation of the probability of occurrence of initial conditions guaranteeing the onset of inflation in the case $X_{\varphi}^{(in)}>0$}

It is noteworthy that there is a fairly wide range of model parameters in which they can be changed without a significant impact on the above qualitative conclusions.
But reasoning in the spirit of chaotic inflation, we can consider $\varphi_{in}$  as a random variable, and then the two above examples with the choice of parameters show a significant difference.  Given that a significant part of the potential is flat, without claiming good accuracy of estimates, we can assume that the probability has a flat prior distribution.
 As can be seen from Eqs.(\ref{eff L general via zeta})-(\ref{K2}),  the TMT effective Lagrangian  and $\zeta(\varphi,X_{\varphi})$  do not change under the replacement
$\varphi\rightarrow -\varphi$. Therefore, we can restrict ourselves to positive $\varphi$ only. Then, with the parameters used in Fig.3, one can find a rough estimate of the probability of occurrence of initial conditions that guarantee the onset  of inflation  as  $(12.9-12.6)/12.9\approx 0.023$. But a similar estimate with the parameters used in Fig.4 gives $(13.7-6.8)/13.7\approx 0.5$, which is a more than optimistic estimate for the probability of occurrence of initial conditions that guarantee the onset of inflation.

\end{itemize}

We must keep in mind that the described  conclusions follow from the analysis of the model (\ref{S without fine tun}) if the integration constant 
$\mathcal{M}$ is chosen so that the cosmological constant is equal to zero (see Eqs.(\ref{L of M}) and (\ref{V1 V2 M})).

\subsubsection
 {The main results in the case $X_{\varphi}^{(in)}<0$}

\hspace{0.7cm} 1) \underline{In the interval $\varphi_{in}<\varphi_0$:}
It is not possible to find such a range of model parameters for which the requirement $\zeta>0$ would be compatible with the condition
$|X_{\varphi}^{(in)}|<V_ {eff}^{ (0)}(\varphi_{in})$. Therefore, we come to the conclusion that, most likely, the initial conditions necessary for the beginning of inflation cannot arise  in the interval $\varphi_{in}<\varphi_0$.

2) \underline{In the infinite interval $\varphi_{in}>\varphi_0$:}
The condition $\zeta>0$ is satisfied in the final interval of $\frac{|X_{\varphi}^{(in)}|}{V_{eff}^{(0)}(\varphi_{in})}$ described by 
the following double inequality  whose ends depend onto $\varphi_{in}$
\begin{equation}
\frac{(2b_p-1)(1+\zeta_v)^2}{k_2(1-b_k)(\zeta_v+b_p)}\left(1-e^{-4(z_{in}-z_0)}\right)<\frac{|X_{\varphi}^{(in)}|}{V_{eff}^{(0)}(\varphi_{in})}<
\frac{(1+\zeta_v)^2}{k_2(\zeta_v+b_p)}\left(\frac{2b_p-1}{\zeta_v}e^{-4(z_{in}-z_0)}+1\right).
\label{double ineq neg X 1}
\end{equation}
This result, which ensures the fulfillment of the condition $\zeta>0$, as we see, can be directly used to check the fulfillment of the initial condition
$|X_{\varphi}^{(in)}|\lesssim V_{eff}^{(0)}(\varphi_{in})$ needed for the onset of  inflation.
With the model parameters used in Fig. 3, for almost all
$\varphi_{in}>\varphi_0$ (with the exception of the values $\varphi_{in}>\varphi_0$ which are very close to $\varphi_0$) {\em the initial condition
$|X_{\varphi}^{(in)}|\lesssim V_{eff}^{(0)}(\varphi_{in})$ needed for the beginning of inflation is incompatible with the  condition $\zeta>0$.}
 {\bf With the model parameters used in Fig. 4,  in the infinite interval $\varphi_{in}>\varphi_0$ the initial condition
$|X_{\varphi}^{(in)}|\lesssim V_{eff}^{(0)}(\varphi_{in})$ required for the beginning of inflation is guaranteed 
 over almost the entire range of restrictions on the values of $X_{\varphi}^{(in)}$  imposed by the inequalities (\ref{double ineq neg X 1}), i.e. by
the condition $\zeta>0$.}  It is only worth noting that  $X_{\varphi}^{(in)}<0$ means the dominance of the spatial gradients of the  field $\varphi(x)$, which usually arise due to quantum fluctuations. The interpretation and possible cosmological effect of this result may be of interest, but its study is beyond the scope of this paper. Leaving this interesting line of research for future work, in Sec.4 we will focus only on the case $X_{\varphi}^{(in)}>0$.

\subsection{About the limit $\zeta=\zeta(\varphi_{in},X_{\varphi}^{(in)}) \rightarrow 0^{\,+}$}

In the next section, we explore the relationship between the conditions for the onset of inflation and what happens in the neighborhood of the  hypersurface
$\zeta(x)=0$. Therefore, it is important to formulate the results of studying what happens when $\zeta\rightarrow 0^{\,+}$. Everything needed for this was done in  Appendix B, and the results for $X_{\varphi}^{(in)}>0$ and $X_{\varphi}^{(in)}<0$ are completely different.

\subsubsection{The case $X_{\varphi}^{(in)}>0$}

{\em If $X_{\varphi}^{(in)}>0$ then  $\zeta\rightarrow 0^{\,+}$ if and only if $\varphi_{in}\rightarrow\varphi_0^{\,-}$.}
 Moreover, as it was shown  in Appendix B (see Sec.B.1, item 1) and Eq.(\ref{phi0 X0 zeta0}) ),
\begin{equation}
\lim_{\varphi_{in} \to \varphi_0^{\,-}}\zeta(\varphi_{in},X_{\varphi}^{(in)})=0 \quad {\text {\bf and}} \quad 
\lim_{\varphi_{in} \to \varphi_0^{\,-}}X_{\varphi}^{(in)}=0.
\label{X>0 lim phi0}
\end{equation}
Thus, for the case $X_{\varphi}^{(in)}>0$ in the plane $(X_{\varphi},\varphi)$ the point $(0,\varphi_0)$ is the boundary point such that
$\varphi_{in}$ cannot be extended to  $\varphi_{in}>\varphi_0$ without changing the sign of $\zeta$ from positive to negative.

\subsubsection{The case $X_{\varphi}^{(in)}<0$}

If $X_{\varphi}^{(in)}<0$, then the constraint written in the form of Eq.(\ref{constr X negative in terms z star}) implies that 
$\zeta(\varphi_{in},X_{\varphi}^{(in)}) \rightarrow 0^{\,+}$ in the infinite interval $\varphi_{in}>\varphi_0$ if and only if
points $(\varphi_{in},|X_{\varphi}^{(in)}|)$ in the half-plane $(\varphi, |X_{\varphi}|)$ tend to the points  $(\bar{\varphi}_{in},\overline{|X_{\varphi}^{(in)}|})$ of the boundary line $l_{(\zeta=0)}$ given by the equation
\begin{eqnarray}
 \overline{|X_{\varphi}^{(in)}|}&=&
\frac{2b_p-1}{1-b_k}\frac{\lambda M_P^4}{4\xi^2}\left(1- e^{-\frac{4\sqrt{\xi}}{M_P}(\bar{\varphi}_{in}-\varphi_0)}\right)
\nonumber
\\
&\approx& V_{eff}^{(0)}(\bar{\varphi}_{in})\frac{(2b_p-1)(1+\zeta_v)^2}{k_2(1-b_k)(\zeta_v+b_p)}
\left(1- e^{-\frac{4\sqrt{\xi}}{M_P}(\bar{\varphi}_{in}-\varphi_0)}\right).
\label{phi bar limit}
\end{eqnarray}
This boundary line $l_{(\zeta=0)}$ starts from some point very close to the point $(\varphi_0,0)$, and as $\bar{\varphi}_{in}>\varphi_0$ increases, the shape of the curve asymptotically quickly approaches the straight line $|X_{\varphi}|=\frac{2b_p-1}{1-b_k}\frac{\lambda M_P^4}{4\xi^2}$ parallel to the $\varphi$ axis. Thus, for the case $X_{\varphi}^{(in)}<0$, in the half-plane $(\varphi,|X_{\varphi}|)$ shifting the point $(\varphi_{in},|X_{\varphi}^{(in)}|)$ for which $\zeta>0$ in the direction of the boundary line $l_{(\zeta=0)}$ it is impossible to cross this line without changing the sign of $\zeta$ from positive to negative. In accordance with the remark made at the very end of Sec.3.2, this result will no longer be discussed in the rest of this paper.

\section{Towards the assertion announced in the title of the paper}

The inclusion of the volume 4-form $dV_{\Upsilon}=\Upsilon(x)d^4x$ in the  least action principle plays a  key role both  in  a possibility of 
the existence of a spacelike boundary surface $\mathcal{B}$ predicted by the BGV theorem and deriving  the constraints on the initial conditions for inflation.
 So far, in order to emphasize that the latter are the results of model dynamics, we have not touched on the intersection of these two issues. Now we have everything we need to: 1) understand that the found dynamical constraints necessary for the onset of inflation entail an initial cosmological singularity in the spirit of the BGV theorem; 2) find conditions on the spacelike boundary surface $\mathcal{B}$ and thus, using the obtained dynamical results, fill the prediction of the BGV theorem with concrete content.

\subsection{Initial coditions for inflation and the spacelike character of the hypersurface $\Upsilon(x)= 0$}

In the models of Secs.2 and 3, after solving Eq.(\ref{varphiB}), we came to a key conclusion: the dynamics of the models dictates that nontrivial solutions exist only when the measure density  $\Upsilon(x)\neq 0$ and, therefore, $\Upsilon(x)$ must be sign-definite. However, after examining the models, we found that the hypersurface $\Upsilon(x)=0$ does indeed exist.
This means that the hypersurface $\Upsilon(x)=0$ separates the space-time manifold $M_4$ into two regions with $\Upsilon(x)>0$ and $\Upsilon(x)<0$, and we must study equations and their solutions in these regions independently of each other. These two regions can be considered as submanifolds  of  ${M_4}$ with a common boundary $\Upsilon(x)=0$. In what follows we will use the notations ${M_4^{(+)}}$ and ${M_4^{(-)}}$ for the submanifolds with $\Upsilon(x)>0$ and $\Upsilon(x)<0$ respectively.

Studying the equations in ${M_4^{(+)}}$, we have chosen an arbitrary integration constant ${\mathcal M}$ to be ${\mathcal M}_0=2\sqrt{V_1V_2}$, Eq.(\ref{V1 V2 M}).  This choice ensures that 1) the vacuum of the classical scalar field $\varphi$ is at $\varphi=0$; 2) the vacuum energy density is zero; 3) the corresponding equations   describe an inflationary model satisfying the Planck's constraint. The zero value of the cosmological constant and possibility of an oscillatory regime of approaching $\varphi$ to its vacuum state allows us to treat this model as the inflationary model of our Universe. Note that by a small deviation from the choice ${\mathcal M}_0$ one can provide a tiny cosmological constant. But this evidently has negligible effect in the inflationary epoch.

As a result of the condition $\Upsilon(x)>0$, or equivalently $\zeta(x)>0$, in Sec.3 and in Appendix B we discovered that, if $X_{\varphi}>0$, there exists an upper bound $\varphi_0$ on admissible values of the inflaton field $\varphi$  for which
\begin{equation}
\Upsilon\equiv\zeta(\varphi,X_{\varphi})\cdot\sqrt{-g}\rightarrow 0^+  \quad \text{and} \quad X_{\varphi}\rightarrow 0^+\quad \text{as} \quad  \varphi\rightarrow\varphi_0^{\,\,\,-}.
\label{upsilon 0}
\end{equation} 
In Sec. 2.4 we have already made an important remark that in the context of the dynamics of the models under consideration, we can describe the hypersurface  $\Upsilon(x)=0$ only in a limiting way, as in Eq. (\ref{upsilon 0}) (or, which is the same, in terms of $\zeta$, as in Eq.(\ref{zin 0}) where $\varphi_*$ should be replaced by 
 $\varphi_0$). If we return to Eq.(\ref{varphiB}) and assume the possibility of $\Upsilon=0$ when solving it, then, at first glance, the remaining equations on this hypersurface simply describe a different dynamics. But it turns out that this notion is erroneous and the reason for this is a new unexpected property on the boundary (the metric tensor has a discontinuity), which will be studied in Sec.4.3.
For the time being, the final conclusion, important for what will be discussed below, is that in the submanifold ${M_4^{(+)}}$ due to the choice of the integration constant ${\mathcal M}_0$ we are dealing with a model of our Universe with the boundary condition (\ref{upsilon 0}). In what follows, for brivety, we will call the latter as "the boundary condition $(\varphi_0;X_{\varphi}=0; [\mathcal {M}_0])$", where $[\mathcal {M}_0]$ is included  to keep in mind that,  in accordance with the constraint (\ref{1+ zeta no fine tun}) and Eq.(\ref{sinh 4 max no fine tuned case}),
the value of $\varphi_0$ is determined by the choice of the integration constant. Recall that the vacuum value $\zeta_v$ present in  (\ref{1+ zeta no fine tun}) and (\ref{sinh 4 max no fine tuned case})  is uniquely determined by the integration constant (see Eqs.(\ref{zeta in vac})-(\ref{zeta in vac 0})). 

Let us now turn to the results of studying  the model of Sec.3, according to which, for the beginning of inflation it is enough
 that   $X_{\varphi}^{(in)}>0$ and the initial value $\varphi_{in}$ of the inflaton field lies in the interval 
$\varphi_{in}^{(min)}\leq\varphi_{in}<\varphi_0$, where $\varphi_{in}^{(min)}$ is defined in Eq.(\ref{interval phi for X positive}).
In addition,  in this  interval of $\varphi_{in}$, the TMT effective  potential has a plateau-like shape and $X_{\varphi}^{(in)}$ is less than the plateau height. 
Therefore, we can assume that the Universe  inflates in the slow-roll regime.

The condition $X_{\varphi}>0$ means that the normal vectors $\partial_{\alpha}\varphi$ 
to all hypersurfaces described by the equations $\varphi(x)=constant$ are timelike and consequently, the hypersurfaces $\varphi(x)=constant$ are spacelike.
All of the above remains true for $\varphi(x)$ arbitrarily close to $\varphi_0$. To find out the geometric properties of the hypersurface $\Upsilon(x)=0$, we must, given the definition $\Upsilon(x)\equiv\zeta(x)\sqrt{-g(x)}$, study what happens to $\zeta(x)\equiv\zeta(\varphi(x),X_{\varphi})$ when $\varphi(x)\to \varphi_0$. In doing this analysis, we must not forget that $\varphi(x)$ and $X_{\varphi}$ are considered independent of each other and serve as initial values for inflation. Therefore, it would be wrong to conclude that $X_{\varphi}\to 0$ because $\varphi(x)\to\varphi_0=const$.
But we can  take into account the very nontrivial result\footnote{For a comparison  with constraints  to the initial values of $\varphi(x)$ and $X_{\varphi}$ if $X_{\varphi}<0$, see Sec.3.3.2.} containing in Eq.(\ref{upsilon 0}):    $X_{\varphi}> 0$  decreases  to 0 when $\varphi(x) \to \varphi_0^{\,\, -}$.
Therefore, for values of $\varphi(x)$ very close to  $\varphi_0$ the function $\zeta(\varphi(x),X_{\varphi})$ can be considered approximately as a function depending only on $\varphi(x)$: $\zeta(\varphi(x),X_{\varphi})\approx \zeta(\varphi(x))$. Hence, for  $\varphi(x)\approx \varphi_0$ 
the equation $\zeta(\varphi(x))=constant$ describes a spacelike  hypersurface. The latter remains true when $\varphi(x) \to \varphi_0^{\,\, -}$ (which is accompanied by $constant \to 0$).
Therefore, the hypersurface  $\Upsilon(x)\equiv\zeta(x)\sqrt{-g(x)}=0$ is a {\bf space-like} {\em boundary of the submanifold  ${M_4^{(+)}}$.}

\subsection{The emergence of a hypersurface $\Upsilon(x)=0$ as an effect of 
\\
spontaneous breaking of the orientability of the space-time manifold $M_4$}

Trying to understand the nature of the  boundary $\Upsilon=0$, we must apparently turn to some basic mathematical definitions and propositions known in the theory of smooth manifolds. It seems that the most appropriate way to understand the problem we are facing is to discuss the orientability of a manifold.

As is well known, the orientation of the differentiable manifold can be defined by the coordinate atlas (collection of maps). But n-dimensional differentiable manifolds (with or without boundary) allow one to define an orientation in an equivalent way using the n-form (see, for example, Chapter 15 in the textbook \cite{Lee}). 
Here it is appropriate to recall the following mathematical proposition, which holds for a differentiable manifolds {\em with and without boundary}, see e.g. Proposition 15.5 in Ref. \cite{Lee}. Applied to $M_4$, the proposition is:  Any smooth nonvanishing   4-form $\omega$ on  $M_4$ determines a unique  orientation  of $M_4$ for which $\omega$ is positively oriented at each point.  Conversely, if an orientation is given on $M_4$, then there is  a smooth nonvanishing   4-form  on  $M_4$ that is positively oriented at each point. 
A similar proposition is true, of course, for a negatively oriented 4-form $\omega$ on $M_4$.
In the TMT model under study, such a 4-form is $dV_{\Upsilon}=\Upsilon d^4x$ defined by Eq.(\ref{Phi}). Thus, {\em  the submanifold $M_4^{(+)}$ (where  $\Upsilon>0$) and  the submanifold $M_4^{(-)}$ (where  $\Upsilon<0$) have opposite spacetime orientations}. Here it is very important to pay attention to the fact that, according to the proposition, the existence of the boundary $\Upsilon(x)= 0$ separating the submanifolds $M_4^{(+)}$ and $M_4^{(-)}$ does not affect this conclusion.

It is worth remembering that by applying the principle of least action to the primordial action (\ref{S without fine tun}), 
we mean, as usual, that all primordial variables (scalar field $\phi$, non-degenerate metric tensor $g_{ \mu\nu}$, 
the affine connection $\Gamma^{\lambda}_{\mu\nu}$, the functions $\varphi_a$ 
and hence $\Upsilon$) are defined globally {\em on an orientable space-time manifold $M_4$} and are smooth functions.
 In particular, this means that: 1) they are all continuous on $M_4$; 2) one can choose the  orientation of $M_4$. 
The latter, according to the  above proposition, should be expressed in the possibility of choosing a certain sign of $\Upsilon(x)$. As we have seen, in order to describe our Universe, when solving equations in the model under consideration, we had to put $\Upsilon(x)>0$, i.e. choose a positive orientation. Based on our experience in field theory, we might naively expect this condition to still hold globally on $M_4$. However, it turns out that this is not the case due to the dynamics of the model. The reason for this phenomenon is a very profound change introduced by TMT into field theory: as a result of the inclusion of the variables $\varphi_a$ (from which $\Upsilon(x)$ is built) into the principle of least action, the possibility arises of {\em mutual influence of the  matter field dynamics and the continuous function $\Upsilon(x)$ or, in other words, between matter and such a fundamental property of the space-time manifold as orientability.} This fundamentally new dynamical effect can lead to the creation of a hypersurface $\Upsilon(x)= 0$ in $M_4$ , which, as we have seen, can in turn impose significant restrictions on the dynamics of mater fields.

The creation of the hypersurface $\Upsilon(x)= 0$ (despite the fact that originally $\Upsilon(x)$ was nonvanishing on $M_4$) and the splitting of $M_4$ into two submanifolds $M_4^{(+)} $ and $M_4^ {(-)}$ with the opposite orientation means that $M_4$ is not oriented anymore. Since this effect is the result of solving dynamical equations, it can be interpreted as {\bf spontaneous violation of the orientability of $M_4$}.

 Moreover, we found that {\em the dynamically arisen spacelike hypersurface $\Upsilon(x)=0$} does not allow extension of solutions for $\varphi(t)$ from $\varphi(t)<\varphi_0$ to $\varphi(t) \geq\varphi_0$.
Note that this effect can also be described in terms of the original scalar field $\phi$, because the replacement (\ref{phi via canonical varphi}) of the original scalar field $\phi$ by the canonically normalized $\varphi$ does not depend on $ \zeta$. Therefore, the origin of this effect has nothing to do with the redefinition (\ref{phi via canonical varphi}) from $\phi$ to $\varphi$.

\subsection{Something about what is hidden behind the space-like hypersurface $\Upsilon(x)=0$
\\
   and boundary conditions on this hypersurface}

When solving Eq.(\ref{varphiB}) in the submanifold $M_4^{(-)}$ (that is in the region with $\Upsilon<0$),
 nothing definite can be said based on what was done in the submanifold $M_4^{(+)}$. The reason is that 
the equations obtained by performing the TMT procedure essentially depend on the choice of the integration
 constant ${\mathcal M}$, which appears in Eq.(\ref{var varphi}) under the condition $\Upsilon(x) \neq 0 $, Eq.(\ref{Phi neq 0}). 
Generally speaking, there is no reason for the values of ${\mathcal M}$ on both sides of the hypersurface $\Upsilon(x)= 0$ to be equal.
After we have implemented in $M_4^{(+)}$ the conditions for the inflationary scenario of our Universe by choosing the integration constant $\mathcal{M}_0=2\sqrt{V_1V_2}$, we would like to have at least some information about possible scenarios in $M_4^{(-)}$. 
Since the integration constant enters into all equations of the model, it is quite expected that the choice of ${\mathcal M}$ can 
radically affect vacuum states and cosmological dynamics in regions with the opposite sign of $\Upsilon$. 
The absence of information about possible scenarios in $M_4^{(-)}$ means that the integration constant in the region with $\Upsilon(x)<0$ can have any real value, and hence the existence of a continuum of different solutions in $M_4^{(-)}$ is possible. Therefore, solutions with the same integration constant $\mathcal{M}_0=2\sqrt{V_1V_2}$ as in the region with $\Upsilon(x)>0$ have zero measure in the space of all possible solutions with $\Upsilon ( x)<0$. That is why we will not consider them further.

 Perhaps more detailed knowledge about  possible cosmological scenarios in $M_4^{(-)}$ is of interest. However, in this paper, we will not engage in additional reasoning in this direction and confine ourselves to a discussion of what happens on the boundary $\Upsilon(x)=0$.
The difference in cosmological dynamics in regions with the opposite sign of $\Upsilon$, in particular, means that one-sided boundary conditions on the hypersurface $\Upsilon(x)=0$ (i.e., in $M_4^{( +)}$ and $M_4^{(-)}$ ) are also different in the general case.
It turns out that the study of this aspect of the model leads to an important conclusion, which we will now deal with.

 In what follows, to avoid confusion, in the submanifold $M_4^{(-)}$ 
we will use the upper symbol 'check' in the notation for the integration constant, the metric in the Einstein frame, the inflaton field
and its derivatives, like  $\check{{\mathcal M}}$,
$\check{\tilde{g}}_{\mu\nu}$, $\check{\varphi}$, $\check{X}_{\check{\varphi}}=
\frac{1}{2}\check{\tilde{g}}^{\alpha\beta}\check{\varphi}_{,\alpha}\check{\varphi}_{,\beta}$.

Let us start by summarizing some of the main results that follow from the general formulation of the model described by Eqs.(\ref{S without fine tun})-(\ref{u zeta model sec 4}) when we apply them to  the submanifold $M_4^{(-)}$, that is in  the region with $\Upsilon<0$.
In the case of an  arbitrary integration constant  $\check{{\mathcal M}}$, the constraint in the Eistein frame in terms of the canonically normalized scalar field 
$\check{\varphi}$ follows from Eq.(\ref{1+ zeta no fine tun phi M}) and, instead of Eq.(\ref{1+ zeta no fine tun}), it has the form as follows
\begin{equation}
\zeta(\check{\varphi},\check{X}_{\check{\varphi}})=\frac{\check{{\mathcal M}}+2|V_1|-
(2b_p-1)\left[\frac{m^2M_P^2}{2\xi}\sinh^2\check{z}+\frac{\lambda}{4\xi^2}M_P^4\sinh^4\check{z}\right]
-(1-b_k)\cosh^4\check{z}\cdot \check{X}_{\check{\varphi}}}
{\check{{\mathcal M}}+2|V_2|+\frac{m^2M_P^2}{2\xi}\sinh^2\check{z}+
\frac{\lambda}{4\xi^2}M_P^4\sinh^4\check{z}+(1-b_k)\cosh^4\check{z}\cdot \check{X}_{\check{\varphi}}}
\label{chec zeta no fine tun M4-}
\end{equation}
where $\check{z}=\sqrt{\xi}\frac{\check{\varphi}}{M_P}$. 
 The transition to the Einstein frame,  Eq.(\ref{gmunuEin}), expressed via the canonically normalised scalar field in $M_4^{(+)}$ is  described by the relation
\begin{equation}
\tilde{g}_{\mu\nu}=[1+\zeta(\varphi,X_{\varphi})]\left(1+\sinh^2\sqrt{\xi}\frac{\varphi}{M_P}\right)g_{\mu\nu}.
 \label{gmunuEin in M4+}
\end{equation}
and in  $M_4^{(-)}$ - by the relation
\begin{equation}
\check{\tilde{g}}_{\mu\nu}=[1+\zeta(\check{\varphi},\check{X}_{\check{\varphi}})]\left(1+\sinh^2\sqrt{\xi}\frac{\check{\varphi}}{M_P}\right)g_{\mu\nu}.
 \label{gmunuEin in M4-}
\end{equation}

As is seen from Eq.(\ref{gmunuEin in M4-}) in order to ensure non-degeneracy of the physical metric tensor $\check{\tilde{g}}_{\mu\nu}$ we have to prevent 
a possibility for $\zeta$ to take the value -1 (see also the important remark made in footnote 4). This means that a set of configurations of the values  $\check{{\mathcal M}}$, $\check{\varphi}$ and $\check{X}_{\check{\varphi}}$ for which the right side of Eq.(\ref{chec zeta no fine tun M4-}) equals -1 has to be excluded from consideration. But it is clear that such a set has zero measure in the set of all possible values of these quantities. Similarly, the denominator in the right  side of 
 Eq.(\ref{chec zeta no fine tun M4-}) can be equal to zero only for very special cases
 which also have zero measure in the set of all possible values of $\check{{\mathcal M}}$, $\check{\varphi}$ and $\check{X}_{\check{\varphi}}$.
Therefore, the scalar $\zeta$ is finite for almost all possible values of these quantities.
Combining the result that $\Upsilon>0$ in $M_4^{(+)}$, the continuity of  $\Upsilon$ which  means that $\Upsilon<0$ in $M_4^{(-)}$, and the definition of $\zeta$, Eq.(\ref{zeta}),  which   in $M_4^{(-)}$ lookes as
\begin{equation}
\zeta(\check{\varphi},\check{X}_{\check{\varphi}})=\frac{\Upsilon}{\sqrt{-g}},
\label{upsilon 0-}
\end{equation}
we obtain that   $\zeta(\check{\varphi},\check{X}_{\check{\varphi}})<0$ and $\zeta\rightarrow 0^{\,-}$ as
$\Upsilon\rightarrow 0^{\,-}$ and vice versa. Taking into account the condition $\zeta\neq -1$, we find the following interval of admissible values of the continuous scalar function $\zeta(x)$ on $M_4^{(-)}$
\begin{equation}
-1<\zeta(\check{\varphi},\check{X}_{\check{\varphi}})<0
\label{zeta interval -}
\end{equation} 

Note that  the structure of the TMT effective action in $M_4^{(-)}$ is the same as in $M_4^{(+)}$ and it is described by Eqs.(\ref{u zeta model sec 4})-(\ref{K2}).   The only difference is that in $M_4^{(+)}$ the  integration constant ${\mathcal M}={\mathcal M}_0=2\sqrt{V_1V_2}=2\zeta_0|V_2|$, while in $M_4^{(-)}$ the  integration constant $\check{{\mathcal M}}\neq {\mathcal M}_0$.
The function $U_{eff}(\check{\varphi},\zeta(\check{\varphi},\check{X}_{\check{\varphi}});\check{{\mathcal M}})$ (from which the  TMT effective potential in $M_4^{(-)}$ is extracted) has the same form as $U_{eff}(\varphi,\zeta(\varphi,X_{\varphi});{\mathcal M})$ in Eq.(\ref{u zeta model sec 4}), where one should replace 
$\varphi$, $\zeta(\varphi,X_{\varphi})$, ${\mathcal M}$ with $\check{\varphi}$, $\zeta(\check{\varphi},\check{X}_{\check{\varphi}})$, $\check{{\mathcal M}}$.
Now we have enough information in order to discuss possible one-sided boundary conditions on the hypersurface $\Upsilon(x)=0$ in $M_4^{(-)}$. 
For an arbitrary $\check{{\mathcal M}}$
 and any finite values of $\check{\varphi}$ and $\check{X}_{\check{\varphi}}$, except for mentioned above very special cases,
the function $U_{eff}(\check{\varphi},\zeta(\check{\varphi},\check{X}_{\check{\varphi}});\check{{\mathcal M}})$ is finite. 
Therefore, for almost  all values of
$\check{{\mathcal M}}$, $\check{\varphi}$ and $\check{X}_{\check{\varphi}}$,
the metric tensor in the Einstein frame $\check{\tilde{g}}_{\mu\nu}$, obtained as a  
solution of the Einstein equations, is regular. 

To demonstrate a qualitatively new result of interest to us, namely, one-sided boundary conditions on the hypersurface $\Upsilon=0$, we now need to study the conditions under which $\Upsilon \rightarrow 0^{\,-}$. This happens if the numerator in
Eq.(\ref{chec zeta no fine tun M4-}) tends to zero as $\check{\varphi}$ tends to the corresponding value $\check{\varphi}_0$. To simplify the analysis, without losing the essence of the new effect, we can assume that $\check{X}_{\check{\varphi}}\to 0$ when $\check{\varphi}\to\check{\varphi}_0 $ (this is similar to the boundary conditions for $\Upsilon\rightarrow 0^{\,+}$ in $M_4^{(+)}$). In addition, for simplicity of calculations, we can assume that $\sqrt{\xi}\frac{\check{\varphi}_0}{M_P}\gg 1$ and neglect the term $\propto \sinh^2\sqrt{\xi}\frac{\check{\varphi}_0}{M_P}$ with respect to the term $\propto \sinh^4\sqrt{\xi}\frac{\check{\varphi}_0}{M_P}$. In this approximation, from Eq.(\ref{chec zeta no fine tun M4-}) we obtain that for a given $\check{{\mathcal M}}$, the limit value $\check{\varphi}_0$ is determined by the relation
\begin{equation}
\sinh^4\sqrt{\xi}\frac{\check{\varphi}_0}{M_P}=\frac{4\xi^2(\check{{\mathcal M}}+2|V_1|)}{\lambda(2b_p-1)M_P^4}.
\label{check varphi 0}
\end{equation} 
Following our agreement we will use $(\check{\varphi}_0;\check{X}_{\check{\varphi}}=0;[\check{{\mathcal M}}])$ for notation of these boundary conditions on the hypersurface $\Upsilon=0$ in $M_4^{(-)}$. In the limit as $\check{\varphi}\rightarrow\check{\varphi}_0$ 
and $\check{X}_{\check{\varphi}}\rightarrow 0$, the function $U_{eff}(\check{\varphi},\zeta(\check{\varphi},\check{X}_{\check{\varphi}});\check{{\mathcal M}})$ tends to the limiting value of the TMT effective potential on the  hypersurface $\Upsilon(x)=0$
\begin{equation}
U_{eff}(\check{\varphi},\zeta(\check{\varphi},\check{X}_{\check{\varphi}});\check{{\mathcal M}})\longrightarrow
\frac{\lambda M_P^4}{4\xi^2}\cdot\frac{b_p\check{{\mathcal M}}+|V_1|}{\check{{\mathcal M}}+2|V_1|}.
\label{u check varphi 0}
\end{equation} 
This limit which is obtained when $\Upsilon(x)\rightarrow 0^{\,-}$ should be compared with the limit  when $\Upsilon(x)\rightarrow 0^{\,+}$.
The latter is controlled by equations of Sec.3.1  when $\varphi\rightarrow\varphi_0^{\,\,\,-}$ (and, according to Eq.(\ref{upsilon 0}), together with  $X_{\varphi}\rightarrow 0^{\,+}$) with the result
\begin{equation}
U_{eff}^{(0)}(\varphi,\zeta(\varphi,X_{\varphi}))\longrightarrow
\frac{\lambda M_P^4(2b_p+\zeta_v)}{8\xi^2(1+\zeta_v)}.
\label{u varphi 0}
\end{equation}
The difference between these two limits, as a matter of fact, describes the difference between the one-sided boundary values of the energy momentum tensor
on the space-like hypersurface $\Upsilon=0$. Therefore, components of the Einstein  tensor $R_{\mu\nu}(\tilde{g})-\frac{1}{2}\tilde{g}_{\mu\nu}R(\tilde{g})$ have jump discontinuity on this hypersurface.
Using the relations describing the transformations to the Einstein frame (\ref{gmunuEin in M4+}) and  (\ref{gmunuEin in M4-}) when  $\zeta\rightarrow 0$, we obtain  the explicit expression for the discontinuity of the metric tensor $\tilde{g}_{\mu\nu}$ on the hypersurface $\Upsilon=0$
\begin{equation}
\lim_{\varphi \to \varphi_0}\tilde{g}_{\mu\nu}-\lim_{\check{\varphi} \to \check{\varphi}_0}\check{\tilde{g}}_{\mu\nu}
=\left[\sinh^2\sqrt{\xi}\frac{\varphi_0}{M_P}-\sinh^2\sqrt{\xi}\frac{\check{\varphi}_0}{M_P}\right]g_{\mu\nu}.
 \label{lim z to 0 gmunuEin in M4+-}
\end{equation}

Thus, we come to the conclusion that an inevitable consequence of the difference in the boundary conditions $(\varphi_0;X_{\varphi}=0; [\mathcal {M}_0])$ and $(\check{\varphi}_0;\check{X}_{\check{\varphi}}=0;[\check{{\mathcal M}}])$ on the hypersurface $\Upsilon=0$ (caused by the difference in the integration constants) is the appearance of {\bf a jump discontinuity of the metric tensor $\tilde{g}_{\mu\nu}$ on the hypersurface $\Upsilon=0$}. But along with the difference in the integration constants,  there is another  theoretical reason for the appearance namely of the metric tensor discontinuity. Indeed, as it follows from (\ref{lim z to 0 gmunuEin in M4+-}),   in the absence of the nonminimal coupling, i.e. in the limit $\xi\rightarrow 0$ the discontinuity of the metric disappears. 
However, since still $\check{{\mathcal M}}\neq {\mathcal M}_0=2\sqrt{V_1V_2}$ and $\check{\varphi}_0\neq \varphi_0$, it follows from Eqs.(\ref{u check varphi 0}) and (\ref{u varphi 0}) that the  jump discontinuity of the curvature  is preserved in the limit $\xi\rightarrow 0$.
  Note that in the limit $\xi\rightarrow 0$ the model leads to a different dynamics, which is also quite regular, as one can see, for example, from
 Eqs.(\ref{1+ zeta no fine tun phi M}) and (\ref{u zeta model sec 4}). 

Against the background of our experience in solving Einstein's equations, the boundary conditions obtained above at first glance look completely unacceptable and unnatural. But we must not forget that, unlike Einstein's GR, which is formulated globally on the entire manifold $M_4$, here we are dealing with conditions on the boundary that separates two (sub)manifolds  $M_4^{(+)}$ and  $M_4^{(-)}$, and on each of them its own Einstein's GR is realized.

\subsection{Initial conditions for inflation
\\
 and the space-like hypersurface $\Upsilon=0$ as a boundary in the BGV theorem}

The results formulated in the previous two subsections allow us to start discussing the main statement mentioned in the title of the paper.
To simplify the analysis, we continue our study within the framework of the single-scalar field  cosmological model, in which {\em our Universe is considered as filling the entire submanifold with boundary $M_4^{(+)}$, where $\Upsilon> 0$ and the boundary is the spacelike hypersurface $\Upsilon=0$}. This means that we completely  ignore the options associated, for example, with a possible scenario of chaotic inflation.

The main assumption is that at the beginning of the expansion of the Universe, the initial value $X_{\varphi}^{(in)}>0$;
recall that $X_{\varphi}$ is given by Eq.(\ref{X as difference}). If, in addition, the initial value  $\varphi_{in}$ of the scalar field lies in the interval
$\varphi_{in}^{(min)}\leq\varphi_{in}<\varphi_0$, then this is sufficient to fulfill  the conditions (\ref{kin and grad less V}) necessary for the onset of inflation. 
The assumption  $X_{\varphi}^{(in)}>0$  was also important  to come to the conclusion that the hypersurface $\Upsilon=0$ is spacelike. Thus, the initial conditions $X_{\varphi}^{(in)}>0$ and $\varphi_{in}^{(min)}\leq\varphi_{in}<\varphi_0$ not only ensure the beginning of inflation, but they are also consistent with the existence of a boundary in the form of the spacelike hypersurface $\Upsilon=0$, due to which it is impossible to shift the initial value of the scalar field $\varphi_{in}$ to the values $\varphi_{in}\geq \varphi_0$. 

The discontinuity of the metric tensor $\tilde{g}_{\mu\nu}$ on the hypersurface $\Upsilon=0$ means that the Christoffel connection coefficients $\{^{\lambda}_{\mu\nu}\}$ of the metric $\tilde{g}_{\mu\nu}$ are not defined  on the hypersurface $\Upsilon=0$.  Consequently, {\bf in  the inflating spacetime of the cosmological model under study,   timelike geodesics  cannot  be extended to the past beyond the spacelike hypersurface $\Upsilon=0$. 
This conclusion agrees with the statement of the BGV theorem, and we have reason to assert that the hypersurface $\Upsilon=0$ is the boundary $\mathcal{B}$, which appears in the formulation of the theorem.}

\subsection{
Possible interpretation of nonorientability}

There may be two alternative reasons for the opposite space-time orientations of
  $M_4^{(+)}$ and $M_4^{(-)}$. First: $M_4^{(+)}$ and $M_4^{(-)}$ have opposite spatial orientations, but the same direction of the arrow of time.
 Second: $M_4^{(+)}$ and $M_4^{(-)}$ have the same spatial orientations, but opposite directions of the arrows of  time. 
Since the submanifolds $M_4^{(+)}$ and $M_4^{(-)}$ are separated by a spacelike hypersurface ${\mathcal B}$, it seems very unnatural (or even impossible ?) to have opposite spatial orientations on both sides of ${\mathcal B}$. But the second interpretation looks much more attractive. Normal vectors to the spacelike hypersurface ${\mathcal B}$ are timelike.  As an example of such a normal vector, consider  $\partial_{\mu}\varphi$:  if $X_{\varphi}>0$ in a neighborhood of ${\mathcal B}$,  then $\partial_{\mu}\varphi$ is a timelike vector. Then the effect of opposite directions of the arrows of  time in $M_4^{(+)}$ and $M_4^{(-)}$ manifests itself in almost opposite directions of the vectors $\partial_{\mu}\varphi$ in $M_4^{(+)}$  and $\partial_{\mu}\check{\varphi}$ in $M_4^{(-)}$. Thus, the appearance of opposite directions of the arrows of time may be a manifestation of the effect of spontaneous violation of the orientability of $M_4$. This geometric interpretation, which is based on the results of the model dynamics, seems to be consistent with a completely different approach based on the idea of opposite directions of thermodynamic arrows of time, proposed in the papers 
\cite{termod arr1}-\cite{Vil arrows}.

Accepting the interpretation that the reason for the opposite space-time orientations in $M_4^{(+)}$ and $M_4^{(-)}$ is the opposite direction of the arrows of time in $M_4^{(+)}$ and $M_4^{(-)}$, the only addition that we can make  is that all processes in $M_4^{(-)}$ proceed in the inverse  direction of time.
 Therefore, the space-like hypersurface $\Upsilon(x)= 0$ is also the boundary for past-directed time-like geodesics in the submanifold $M_4^{(-)}$, that is, $M_4^{(-)}$ is also past-geodesically incomplete.

\section{Discussion}

In the framework of TMT in the Palatini formalism, we explored the  scalar field model with the potential 
$V(\phi)=\frac{1}{2}m^2\phi^2+\frac{\lambda}{4}\phi^4$ and non-minimal coupling to gravity.
The inclusion of the volume 4-form $\Upsilon d^4x$ (defined by Eq.(\ref{Phi})) along with the standard Riemannian 4-form $\sqrt{-g}d^4x$ in the principle of least action has far-reaching consequences. Some of them, studied in detail in Secs.3 and 4, look completely unexpected against the background of our experience  in field theory with only the Riemannian volume form. Under certain conditions, the possibility of mutual influence of the dynamics of matter fields and the continuous function $\Upsilon(x)$ arises, and  it can manifest itself as a mutual influence between matter and such a fundamental property of the space-time manifold as orientation. As we have seen, in the model under study this mutual influence turns out to be so significant that the dynamics of the scalar field is capable of causing spontaneous violation of the orientability of the space-time manifold.

For a more accurate understanding of the novelty introduced into the field theory by the volume 4-form $\Upsilon d^4x$, it is useful to compare it with the gravitational action of matter (in the model under consideration, this is the scalar field $\varphi(x)$). In the simplest model of Sec.2, as seen from the constraint (\ref{zeta fine tun}), $\zeta(x)$ is a local function of $\varphi(x)$: $\zeta(x)=\zeta(\varphi(x))$. Conversely, $\zeta(x)$ also enters the $\varphi$ field equation and the energy-momentum tensor as a local function $\zeta(\varphi(x))$. On the contrary, in general relativity the influence of matter on the metric tensor is found as a solution to the Einstein equations, and therefore the metric tensor in the Einstein frame ${\tilde g}_{\mu\nu}(x)$ is an integral over the space-time distribution of matter.
In the  model of Sec.3, this comparison becomes more complicated, since now  the constraint (\ref{1+ zeta no fine tun phi M}) describes $\zeta(x)$ as a local function of two variables: $\zeta(x) =\zeta(\varphi(x),X_{\varphi}(x))$, where $X_{\varphi}=\frac{1}{2}\tilde{g}^{\alpha\beta}\varphi_{,\alpha}\varphi_{,\beta}$ depends on ${\tilde g}_{\mu\nu}(x)$, which in a self-consistent problem must be found as a solution to the Einstein equations.
Therefore, in the general case, $\zeta(x)$ can be affected by matter not only locally. However, in this paper, our attention has been focused on what happens near the  hypersurface $\Upsilon =0$, that is, when $\Upsilon\equiv\zeta(\varphi,X_{\varphi})\cdot\sqrt{-g}\rightarrow 0^+$.  This happens as 
$\varphi\rightarrow\varphi_0^{\,\,\,-}$ and then $X_{\varphi}\rightarrow 0^+$ (see Eq.(\ref{upsilon 0})). 
Therefore, such a strong effect as the spontaneous violation of the orientability of the space-time manifold is in fact a consequence of {\em the \underline{local} nature of the mutual influence of the dynamics of the scalar field $\varphi(x)$ and the density $\Upsilon(x)$ of the volume measure $dV_{\Upsilon}$ on the spacelike hypersurface $\Upsilon(x) =0$}.

Starting with the problem of initial conditions for inflation and assuming that $\rho_{kin}^{(in)}>\rho_{grad}^{(in)}$, we discovered the existence of a spacelike hypersurface $\Upsilon(x)=0$. The boundary conditions on the hypersurface 
$\Upsilon(x)=0$, which are an unambiguous consequence of the model dynamics, are the cause  of the past  directed timelike  geodesic incompleteness of the inflationary  spacetime. This allows us to identify the spacelike hypersurface $\Upsilon(x)=0$ with the boundary $\mathcal{B}$, the existence of which is predicted by the BGV theorem,

Due to the presence of the volume 4-form $\Upsilon(x)d^4x$ in the principle of least action, a new possibility (or even a logical necessity) opens up for adding one more vacuum-like term $\propto \frac{\Upsilon^2}{\sqrt{-g}}V_2$ to the primordial Lagrangian in addition to the usual $\propto\sqrt{-g}V_1$. It turns out that this novelty is decisive for achieving most of the results obtained. This is easy to verify: in the limit $V_2\rightarrow 0$, almost all physical quantities tend to zero or to infinity

Finally, it is worth noting that if we really believe in the efficiency of mathematics, then we have no reason to ignore the possibility of including the 4-form $\Upsilon d^4x$ as a volume measure in the primordial action. Moreover, this seems quite natural, since the construction of smooth manifolds, and in particular a four-dimensional space-time manifold, begins with equipping the topological space with a differentiable structure. As a result, the volume 4-form appears even before the 4D differentiable manifold is endowed with an affine connection and a metric \cite{Lee},\cite{Hawking}. In this regard, the main conclusion that follows from this paper is that the scene on which we build our cosmological models can be significantly expanded if we recognize that orientability is not a self-evident property of the space-time manifold.

\appendix

\section{Two volume elements. 
\\
Brief description of the procedure  of Two-Measure Theory}
\label{TMT procedure}

The main idea of TMT is that in the four-dimensional space-time, in addition to the terms included in the primordial action of the theory with the volume element $dV_g=\sqrt{-g}d^4x$, there are terms that appear in the primordial action with a metric-independent volume element
$dV_{\Upsilon}$ defined by Eq.(\ref{Phi}).
The use of the  volume element $dV_{\Upsilon}$ in the action integral along with $dV_g$ leads to very intersting results studied
 in the series of papers  (see e.g.  \cite{GK1} - \cite{EINRosen}).  Here we formulate the main ideas of TMT and describe the algorithm in general terms, listing all the steps to obtain the results of the theory.
The primordial action  in TMT  in the Palatini formalism includes the following set of variables: 4 functions $\varphi_{a}$ from which $\Upsilon$ is built, 
affine connection $\Gamma^{\lambda}_{\mu\nu}$, primordial (original) metric tensor $g_{\mu\nu}$, and matter fields. All these variables are assumed to be independent, 
and the relationship between them is the result of applying the least action principle. 
As far as the relationship between connection and metric is concerned, this is reminiscent of Palatini's formalism,  which are well studied in models of the usual theory containing only the volume element $dV_g$.
However, the inclusion of one more volume element, 
$dV_{\Upsilon}$, significantly expands the consequences of the Palatini formalism.

The TMT procedure  consists of the following steps:
\begin{itemize}

\item

1. Variation with respect to scalar functions $\varphi_a$ ($a=1,..,4$) yields an equation  solution of which exists if $\Upsilon\neq 0$ and it contains an integration constant ${\mathcal M}$ of dimentionality $(mass)^4$.

\item

2. 
Variation with respect to the affine connection leads to a differential equation, the solution of which gives the connection coefficients. Besides the original metric and its derivatives, this solution for the connection coefficients may also contain matter fields and their first derivatives. In the context of TMT, when applying the Palatini formalism, a new aspect arises: namely, the differential equation for the connection coefficients and its solution also include the 
 scalar $\zeta(x)$ and  its first derivatives (the  scalar $\zeta(x)$ is defined by Eq.(\ref{zeta}).
Obviously, the connection coefficients thus obtained are non-Riemannian (the covariant derivative of the primordial metric tensor with such connection coefficients is nonzero).

\item

3. Varying  the primordial metric results in the equations similar to what occurs in the usual Palatini formalism, but now they also contain the scalar $\zeta(x)$.

\item

4. The condition for the compatibility of the equations obtained at steps 1 and 3 has the form of an algebraic equation describing the scalar $\zeta(x)$ as a local function of matter fields; it also contains an arbitrary integration constant ${\mathcal M}$.
 In Refs.\cite{GK1}-\cite{GK6}, this equation is called 'constraint', which we  also use in this paper. 

\item

5. All matter field equations obtained by variation of the primordial action contain the scalar $\zeta(x)$ and its derivatives.

\item

6. By a  Weyl transformation of the primordial metric $g_{\mu\nu}$, one can simultaneously ensure that the new metric $\tilde{g}_{\mu\nu}$ is Riemannian, and the connection coefficients found in step 2 are converted into the Christoffel coefficients of  $\tilde{g}_{\mu\nu}$.   As a result, the equations obtained in step 3 take exactly the form of  Einstein's equations with the same Newton constant. Therefore $\tilde{g}_{\mu\nu}$ is the metric in the Einstein frame. The energy-momentum tensor arising in the Einstein frame, which we call 'the TMT effective energy-momentum tensor',
contains the scalar $\zeta(x)$, which, in turn, due to the constraint mentioned in step 4, is expressed in terms of matter fields. The potential of scalar field entering to the TMT effective energy-momentum tensor will be called 'the TMT effective potential'. A similar thing happens in the matter field equations. This feature of TMT is a source of very interesting and fundamentally new effects, which will also be demonstrated in this paper.

\item

7. The sequential execution of the steps described above will be referred to as the "TMT procedure". Finally, to check if the TMT procedure was performed correctly, as well as for the convenience of working with the obtained equations in the Einstein frame, one can construct an action whose variation leads to these equations. For this action, we will use the term 'TMT effective action'. The terms 'TMT effective energy-momentum tensor',  'TMT effective potential' and 'TMT effective action' are introduced in order not to confuse them with similar terms used when quantum corrections are taken into account.

\end{itemize}

In order to avoid misunderstanding and confusion regarding {\em the methods of obtaining model results}, it seems necessary to state {\em the difference between TMT models and "conventional" alternative theories of gravity} (that is, models with only one standard Riemannian volume measure). {\em This difference consists only in the order of operations:}  in the conventional alternative models, the transition to the Einstein frame in the original action is first performed, after which, by varying the resulting action, equations are obtained; in TMT models, first, equations are obtained by varying the original (primordial) action, and then a transition to the Einstein frame is performed in these equations.
We also note that if the procedure used in conventional  alternative gravity theories were carried out in TMT, then this would lead to the disappearance of $\Upsilon$ in the volume measure, at least in part of the terms, and its appearance in the Lagrangian, which would lead to completely different theory.

\section{Analysis of the condition $\zeta>0$ in the model of Sec.3} 

\subsection
{Constraint on the maximum initial value of $X_{\varphi}>0$ in a model with $b_k\neq 1$ and $b_p\neq 1$}

The results obtained in this Appendix follow from a detailed study of the inequality (\ref{1+ zeta no fine tun larger 0}).
It follows from the discussion in Sec.3.2 
that the sign of initial $X_{\varphi}$, $X_{\varphi}^{(in)}$, can significantly affect the results.
 $X_{\varphi}^{(in)}$ can be positive or negative depending on how inhomogeneous and anisotropic at the beginning of inflation was the space domain  whose expansion generates our Universe.
In this subsection, the analysis is carried out under the assumption that $X_{\varphi}^{(in)}>0$. In the next subsection, an analysis is made for the case $X_{\varphi}^{(in)}<0$.

To fulfill the requirement $\zeta>0$ for $X_{\varphi}\geq 0$, the values of $\varphi$ and $X_{\varphi}$ must satisfy the inequality
\begin{equation}
2|V_2|\zeta_v(1+\zeta_v)-(2b_p-1)\left[\frac{m^2M_P^2}{2\xi}\sinh^2z+\frac{\lambda}{4\xi^2}M_P^4\sinh^4z\right]
-(1-b_k)\cosh^4z\cdot X_{\varphi}>0
\label{1+ zeta no fine tun larger 0 X positive}
\end{equation}
which follows from the inequality (\ref{1+ zeta no fine tun larger 0}). In particular, the inequality must be true for initial values   $\varphi_{in}$ and $X_{\varphi}^{(in)}$. 
Given that we are interested in the range $z_{in}=\sqrt{\xi}\frac{\varphi_{in}}{M_P}\gg 1$, the inequality can be represented as the following upper constraint on $X_{\varphi}^{(in)}$ which depends on $\varphi_{in}$
\begin{equation}
0\leq\frac{X_{\varphi}^{(in)}}{M_P^4}< \frac{1}{1-b_k}\left[32q^4\zeta_v(1+\zeta_v)e^{-4\sqrt{\xi}\frac{\varphi_{in}}{M_P}}-4(2b_p-1)\frac{\lambda}{4\xi^2}\right]
\label{restr on X from above}
\end{equation}

It is convenient to describe the position of $\varphi_{in}$ relative to $\varphi_0$  defined by Eq.(\ref{sinh 4 max no fine tuned case}).
Since $\sqrt{\xi}\frac{\varphi_0}{M_P}\gg 1$, Eq.(\ref{sinh 4 max no fine tuned case}) reduces to
\begin{equation}
 32q^4\zeta_v(1+\zeta_v)=(2b_p-1)\frac{\lambda}{4\xi^2}e^{4\sqrt{\xi}\frac{\varphi_0}{M_P}}.
\label{phi star}
\end{equation}
Inserting (\ref{phi star}) to Eq.(\ref{restr on X from above}) we obtain 
\begin{equation}
0\leq X_{\varphi}^{(in)}< \frac{\lambda (2b_p-1)}{4\xi^2(1-b_k)}M_P^4\left[\exp\left(4\sqrt{\xi}\left(\frac{\varphi_0}{M_P}-
\frac{\varphi_{in}}{M_P}\right)\right)-1\right]\stackrel{\mathrm{def}}{=}\widehat{X(\varphi_{in})},
\label{restr on X from above final}
\end{equation}
where {\bf $\widehat{X(\varphi_{in})}$  describes the  limiting upper bound of admissible  values of  $X_{\varphi}^{(in)}\geq 0$ as a function of $\varphi_{in}$}. More precisely, here {\em  the term "limiting upper bound of admissible  values of  $X_{\varphi}^{(in)}\geq 0$" means that $\zeta>0$  for any
$X_{\varphi}^{(in)}$ satisfying 
$0\leq X_{\varphi}^{(in)}<\widehat{X(\varphi_{in})}$ and $\zeta<0$ for any $X_{\varphi}^{(in)}>\widehat{X(\varphi_{in})}$}. 
Note that $\widehat{X(\varphi_{in})}>0$ if $\varphi_{in}<\varphi_0$. We see that: 

1) When $\varphi_{in}\rightarrow \varphi_0^{\,\, -}$ we get that $\widehat{X(\varphi_{in})}$ and hence $X_{\varphi}^{(in)}$ and $\zeta=\zeta(\varphi^{(in)},X_{\varphi}^{(in)})$ tend to zero while remaining positive
\begin{equation}
\lim_{\varphi \to \varphi_0^{\,\, -}}\widehat{X(\varphi_{in})}=\lim_{\varphi \to \varphi_0^{\,\, -}}X_{\varphi}^{(in)}=0 \quad \text{{\bf and}} \quad \lim_{\varphi \to \varphi_0^{\,\, -}}\zeta(\varphi_{in},X_{\varphi}^{(in)})=0,
\label{phi0 X0 zeta0}
\end{equation}
which will be especially important in Sec.4, where we  study  what happens near the hypersurface $\Upsilon(x)=0$.

2) Just as it was in the model of Sec.2, where
$\zeta$ does not depend on $X(\varphi)$, {\em the condition $\zeta>0$ forbids $\varphi_{in}$ to be $\varphi_{in}\geq\varphi_0$.}  It is important to recall that we come to this conclusion under the condition $X_{\varphi}^{(in)}>0$. In the case of $X_{\varphi}^{(in)}<0$ this statement is false (see subsection B.2 of this Appendix).

 3) $\widehat{X(\varphi_{in})}$ increases monotonically  as $\varphi_{in}$ decreases.
This means that by shifting $\varphi_{in}$ in a decreasing direction, one can reach a state $\varphi_{in}^{(min)}$, where $\widehat{X(\varphi_{in})}$ becomes equal to $V_{eff}^{(0)}(\varphi_{in})$, and for $\varphi_{in}<\varphi_{in}^{(min)}$ we get 
$\widehat{X(\varphi_{in})}>V_{eff}^{(0)}(\varphi_{in})$.
 Thus, {\em in the interval  $\varphi_{in}^{(min)}<\varphi_{in}<\varphi_0$  the condition $X_{\varphi}^{(in)}<V_{eff}^{(0)}(\varphi_{in})$ is satisfied automatically}.
 The value of $\varphi_{in}^{(min)}$ can be estimated by  the approximate equality
\begin{eqnarray}
\widehat{X(\varphi_{in})}|_{max}&=&\widehat{X(\varphi_{in})}|_{\varphi_{in}=\varphi_{in}^{(min)}}=
 \frac{\lambda (2b_p-1)}{4\xi^2(1-b_k)}M_P^4\left[\exp\left(4\sqrt{\xi}\left(\frac{\varphi_0}{M_P}-\frac{\varphi_{in}^{(min)}}{M_P}\right)\right)-1\right]
\nonumber
\\
&&\approx V_{eff}^{(0)}(\varphi_{in}^{(min)})\approx \frac{k_1\lambda (\zeta_v+b_p)}{4\xi^2(1+\zeta_v)^2}M_P^4,
\label{approx X to hight}
\end{eqnarray}
where the dimensionless factor $k_1\geq 1$ is introduced in order to take into account the change in the height of the potential  when 
$\varphi_{in}$ changes on the segment where the first plateau passes into the second plateau. In the case presented  in Fig.3, we see that $1\leq k_1\lesssim 1.1$, whereas for the shape of $V_{eff}^{(0)}(\varphi)$ in Fig.4 we can take $1\leq k_1\lesssim 1.03$. Since we are interested in the very existence of the interval
of $\varphi_{in}$, in which the conditions for inflation are being prepared, we can restrict ourselves to a qualitative account of this minor change.
Thus,  it follows from Eqs.(\ref{restr on X from above final})-(\ref{approx X to hight}) that {\bf if $X_{\varphi}^{(in)}\geq 0$ and the initial value $\varphi_{in}$ of the canonically normalized classical inflaton field is in the interval}
\begin{equation}
\varphi_{in}^{(min)}\leq  \varphi_{in}
<\varphi_0, \quad \text{where} \quad\varphi_{in}^{(min)}=\varphi_0-  \frac{M_P}{4\sqrt{\xi}}\ln\left(1+\frac{k_1(1-b_k)(\zeta_v+b_p)}{(2b_p-1)(1+\zeta_v)^2}\right),
\label{interval phi for X positive}
\end{equation}
{\bf then both the condition $\zeta>0$ and the condition $X(\varphi_{in})<V_{eff}^{(0)}(\varphi_{in})$ required for the onset of inflation are guaranteed.}

Although for $\varphi_{in}<\varphi_{in }^{(min )}$, as we found out,  $\widehat{X(\varphi_{in})}>V_{eff}^{(0)}(\varphi_{in})$ , the value of $X_{\varphi}^{(in)}$ may still satisfy the  condition $X_{\varphi}^{(in)}<V_{eff}^{(0)}(\varphi_{in})$ required for the onset of inflation.
And if $X_{\varphi}^{(in)}<\widehat{X(\varphi_{in})}$, then $\zeta$ remains positive. Therefore, the conditions necessary for the beginning of inflation are also possible for $\varphi_{in}<\varphi_{in}^{(min)}$, but their fulfillment in this interval is not guaranteed.

\subsection{Constraints on the  initial values of $|X_{\varphi}|$ in a model with $b_k\neq 1$ and $b_p\neq 1$ if $X_{\varphi}<0$}

If $X_{\varphi}^{(in)}<0$, i.e. the initial gradient energy density is larger than the initial kinetic energy density, then it is useful to rewrite the constraint (\ref{1+ zeta no fine tun}) in the form
\begin{equation}
\zeta=\zeta(\varphi_{in},X_{\varphi}^{(in)})=\frac{32q^4\zeta_v(1+\zeta_v)-(2b_p-1)\frac{\lambda}{4\xi^2}e^{4z_{in}}+(1-b_k)\frac{|X_{\varphi}^{(in)}|}{M_P^4}e^{4z_{in}}}
{32q^4(1+\zeta_v)+\frac{\lambda}{4\xi^2}e^{4z_{in}}-(1-b_k)\frac{|X_{\varphi}^{(in)}|}{M_P^4}e^{4z_{in}}},
\label{constr X negative}
\end{equation}
where  $z_{in}=\sqrt{\xi}\frac{\varphi_{in}}{M_P}\gg 1$
and, as before, the negligible terms $\propto e^{2z_{in}}$ are omitted. Again, as in the previous subsection, it is  convenient to describe the position of 
$\varphi_{in}$ relative to $\varphi_0$  defined by Eq.(\ref{sinh 4 max no fine tuned case}). Then it follows from Eq.(\ref{constr X negative}) that the condition $\zeta>0$ is sutisfied if 
\begin{equation}
\zeta=\frac{(2b_p-1)\frac{\lambda}{4\xi^2}\left(e^{4(z_0-z_{in})}-1\right)+(1-b_k)\frac{|X_{\varphi}^{(in)}|}{M_P^4}}
{\frac{\lambda}{4\xi^2}\left(\frac{2b_p-1}{\zeta_v}e^{4(z_0-z_{in})}+1\right)-(1-b_k)\frac{|X_{\varphi}^{(in)}|}{M_P^4}}\equiv \frac{N}{D}>0,
\quad \text{where} \quad z_0=\sqrt{\xi}\frac{\varphi_0}{M_P}
\label{constr X negative in terms z star}
\end{equation}

It turns out that the analysis of restrictions on $\varphi_{in}$ that ensure both $\zeta>0$ and
 the condition $|X_{\varphi}^{(in)}|\lesssim V_{eff}^{(0 )}(\varphi_{in})$ should be carried out separately for the cases $\varphi_{in}<\varphi_0$ and $\varphi_{in}>\varphi_0$. Note that in the case $X_{\varphi}<0$ the condition $|X_{\varphi}^{(in)}|\lesssim V_{eff}^{(0 )}(\varphi_{in}) $ implies the fulfillment of the conditions (\ref{kin and grad less V}) nesessery for the onset of inflation.

\subsubsection{The case of the interval $\varphi_{in}<\varphi_0$}

Obviously, in this case the numerator $N$ in (\ref{constr X negative in terms z star}) is positive. The denominator D$ >0$ if

\begin{equation}
|X_{\varphi}^{(in)}|< \frac{\lambda}{4\xi^2(1-b_k)}M_P^4\left[\frac{2b_p-1}{\zeta_v}\exp\left(4\sqrt{\xi}\left(\frac{\varphi_0}{M_P}-
\frac{\varphi_{in}}{M_P}\right)\right)+1\right]
\stackrel{\mathrm{def}}{=}\widehat{|X(\varphi_{in})|}.
\label{restr on |X| case 1}
\end{equation}
where $\widehat{|X(\varphi_{in})|}$  describes the  limiting upper bound  of admissible  values of  $|X_{\varphi}^{(in)}|$ as a function of $\varphi_{in}$.
Trying to repeat the analysis similar to that which was made after  Eq.(\ref{restr on X from above final}), we also notice that
$\widehat{|X(\varphi_{in})|}$
decreases monotonically 
as $\varphi_{in}$ increases.
 But unlike the case when $X_{\varphi}>0$, now limit of
 $\widehat{|X(\varphi_{in})|}$  as $\varphi_{in}\rightarrow \varphi_0^{\,\,\,-}$ is non-zero in general:
\begin{equation}
\lim_{\varphi \to \varphi_0^{\,\, -}}\widehat{|X(\varphi_{in})|}=  \frac{\lambda}{4\xi^2(1-b_k)}\left(\frac{2b_p-1}{\zeta_v}+1\right)M_P^4
\label{|X| tends nonzero}
\end{equation}
In the case of the parameters used in Fig.3, this limit is equal to
$2\cdot 10^{-10}M_P^4$, and for the parametrs used in Fig.4 this limit is $\approx 4.5\cdot 10^{-10}M_P^4$. 
Despite the significant difference in the shape of the effective TMT potential in these two examples, we find that the minimum value of 
$\widehat{|X(\varphi_{in})|}$ exceeds the height of the potential plateau given by Planck's data. By numerical verification in a wide range of model parameters, we failed to find the possibility that $\widehat{|X(\varphi_{in})|}$ does not exceed $10^{-10}M_P^4$ in a significant way.
This allows us to conclude that in the case of $X_{\varphi}<0$,  most likely, the initial conditions necessary for the beginning of inflation cannot arise  in the interval $\varphi_{in}<\varphi_0$.

\subsubsection{The case of the interval $\varphi_{in}>\varphi_0$}

In this case, to ensure  $\zeta>0$, we have to consider two options: (A) $N>0$, $D>0$; (B) $N<0$, $D<0$.

Let us start from the option (B). From Eq.(\ref{constr X negative in terms z star}), as a result of combyning the conditions  $N<0$ and $D<0$, we get a double inequality
\begin{equation}
\frac{\lambda}{4\xi^2}\left(1+\frac{2b_p-1}{\zeta_v}e^{-4(z_{in}-z_0)}\right)  < \frac{1-b_k}{M_P^4}|X_{\varphi}^{(in)}|<
(2b_p-1)\frac{\lambda}{4\xi^2}\left(1-e^{-4(z_{in}-z_0)}\right),
\label{constr X negative 2A}
\end{equation}
which implies that
\begin{equation}
(2b_p-1)\left(\frac{1}{\zeta_v}+1\right)e^{-4(z_{in}-z_0)}<2(b_p-1).
\label{constr X negative 2A imposs}
\end{equation}
But for any $0.5<b_p<1$ the latter is impossible, which means that in the case of option (B)  $\zeta$ cannot be positive.

Now let us consider the option (A).

For $\varphi_{in}>\varphi_0$ the numerator $N$ in the inequality (\ref{constr X negative in terms z star}) is positive if
\begin{equation}
(1-b_k)\frac{|X_{\varphi}^{(in)}|}{M_P^4}>
(2b_p-1)\frac{\lambda}{4\xi^2}\left(1-e^{-4(z_{in}-z_0)}\right).
\label{constr X negative N positive}
\end{equation}
wheares the denominator $D$ is positive if
\begin{equation}
(1-b_k)\frac{|X_{\varphi}^{(in)}|}{M_P^4}<
\frac{\lambda}{4\xi^2}\left(\frac{2b_p-1}{\zeta_v}e^{-4(z_{in}-z_0)}+1\right).
\label{constr X negative D positive}
\end{equation}
Combining (\ref{constr X negative N positive}) and (\ref{constr X negative D positive}) we get that $\zeta>0$ for any $\varphi_{in}>\varphi_0$ if
\begin{equation}
(2b_p-1)\left(\frac{1}{\zeta_v}+1 \right)e^{-4(z_{in}-z_0)}+2(1-b_p)>0
\label{X negative ND cons}
\end{equation}
and if   $X_{\varphi}^{(in)}<0$ satisfies the inequality (\ref{constr X negative N positive}). Note that the inequality (\ref{X negative ND cons}) is satisfied by our choice of $0.5<b_p<1$. Thus, contrary to what we learned earlier, i.e. when $X_{\varphi}^{(in)}> 0$, if the gradient energy density prevails over the kinetic energy density, then the range $\varphi_{in}>\varphi_0$ may  not be an artifact  and this becomes possible due to $X_{\varphi}^{(in)}<0$. However, inflationary expansion in this range of initial values $\varphi_{in}$ and $X_{\varphi}^{(in)}$ can only start if the upper bound  on 
$|X_{ \varphi}^{(in)}|$ satisfies inequality
\begin{equation}
|X_{\varphi}^{(in)}|\lesssim V_{eff}^{(0)}(\varphi_{in})\approx \frac{k_2\lambda (\zeta_v+b_p)}{4\xi^2(1+\zeta_v)^2}M_P^4,
\label{constr X negative cond for dSitt}
\end{equation}
which is {\em an additional requirement here, in contrast  to what was in the case $X_{\varphi}^{(in)}>0$ considered in  Appendix B.1}.
Here the factor $1<k_2\lesssim 1.3$ is introduced because
taking into account the shape of the TMT effective potential $V_{eff}^{(0)}(\varphi)$ in the case shown in Fig.3, we see that for $\varphi_{in}>\varphi_0$ the scenario is not ruled out when inflation occurs in two stages: first through a higher plateau, and then through a lower plateau. Interestingly, in the case shown in Fig.4, $\varphi_0$ is the value that already corresponds to the second plateau (note that, compared to Fig.3, the presence of the second plateau in Fig. 4 is much less obvious) .

Combyning  (\ref{constr X negative N positive}) and  (\ref{constr X negative D positive}) and using (\ref{constr X negative cond for dSitt}) we get the following double inequality
\begin{equation}
\frac{(2b_p-1)(1+\zeta_v)^2}{k_2(1-b_k)(\zeta_v+b_p)}\left(1-e^{-4(z_{in}-z_0)}\right)<\frac{|X_{\varphi}^{(in)}|}{V_{eff}^{(0)}(\varphi_{in})}<
\frac{(1+\zeta_v)^2}{k_2(\zeta_v+b_p)}\left(\frac{2b_p-1}{\zeta_v}e^{-4(z_{in}-z_0)}+1\right),
\label{double ineq neg X}
\end{equation}
which describes   the condition $\zeta>0$ together with the additional requirement (\ref{constr X negative cond for dSitt}) needed for the onset of inflation.
Consider separately two examples shown in Fig.3 and Fig.4.

\underline{In the case of the model parameters used in Fig.3}, if $k_2=1.3$ is used,
 (\ref{double ineq neg X}) takes the form 
\begin{equation}
1.64\left(1-e^{-4(z_{in}-z_0)}\right)<\frac{|X_{\varphi}^{(in)}|}{V_{eff}^{(0)}(\varphi_{in})}<
1.23\left(2e^{-4(z_{in}-z_0)}+1\right),
\label{double ineq neg X Fig 3}
\end{equation}
From this double inequality follows the conclusion: if $X_{\varphi}^{(in)}<0$, then with the model parameters used in Fig.3, for almost all
$\varphi_{in}>\varphi_0$ (with the exception of $\varphi_{in}>\varphi_0$ which are very close to $\varphi_0$) {\em the initial condition 
(\ref{constr X negative cond for dSitt}) needed for the beginning of inflation is incompatible with the  condition $\zeta>0$.}

\underline{With  the model parameters used in Fig.4},  (\ref{double ineq neg X}) takes the form 
\begin{equation}
8\cdot 10^{-6}\cdot\left(1-e^{-4(z_{in}-z_0)}\right)<\frac{|X_{\varphi}^{(in)}|}{V_{eff}^{(0)}(\varphi_{in})}<
2\left(10^{-5\cdot}e^{-4(z_{in}-z_0)}+1\right),
\label{double ineq neg X Fig 4}
\end{equation}
from which we conclude that if $X_{\varphi}^{(in)}<0$, then for any $\varphi_{in}>\varphi_0$ the joint fulfillment of the condition $\zeta>0$ and the initial condition (\ref{constr X negative cond for dSitt})  necessary for the beginning of inflation is possible. Moreover, since the right side of the inequality (\ref{double ineq neg X Fig 4}) is only 2 times greater than the value necessary as a condition for the onset of inflation, we come to the conclusion that {\em almost in the entire range of restrictions on the values of $X_{\varphi}^{(in)}$  imposed by the condition $\zeta>0$, in the infinite interval $\varphi_{in}>\varphi_0$, the condition (\ref{constr X negative cond for dSitt})  necessary for the beginning of inflation is guaranteed.}


\section{About quantum corrections}

The effect of quantum corrections to the model under study and the related possible deviations from the plateau-like form of the TMT effective potential are of particular interest. 
An analysis of possible approaches to solving the problem of calculating quantum corrections in the TMT models  shows that TMT dictates the need to perform calculations based on the following algorithm.

1) Quantum corrections must be calculated not with the use of the TMT effective action, but with the primordial action. Indeed, for example, the TMT effective potential includes a contribution from the scalar $\zeta$, the value of which is determined from the constraint in the zeroth approximation. Calculating the quantum effective potential in the usual way using the TMT effective potential as the zeroth approximation, we still take into account $\zeta$ in the zeroth approximation. Thus, we ignore the quantum corrections to the constraint, and hence to the $\zeta$. Compare this with the approach in which we start by calculating the quantum corrections to the primordial action. Then, proceeding, for example, from the effective primordial action calculated in the one-loop approximation, following the standard TMT procedure, we can obtain both the constraint and the effective TMT action, in which one-loop corrections are taken into account.
 
2) To calculate the quantum corrections to the primordial action (\ref{S without fine tun}), one can apply the standard technique for calculating the effective potential in the background gravitational field. But in TMT this stage is complicated by the presence of the volume measure $\Upsilon d^4x$ in the primordial action. By redefining some parameters and variables, the contribution of $\Upsilon=\zeta \sqrt{-g}$ can be absorbed so that the primordial action is reduced to the standard form which is considered as the action for a scalar field in the background gravitational field. When making the above redifinitions, we must treat the scalar $\zeta$ as another constant background field. 

3) After finding the one-loop effective action for a scalar field in the background gravitational field using the standard technique, we must return to the original parameters and variables using inverse redefinitions. 
As a result, we obtain the TMT action in the one-loop approximation, in which again there are terms with the standard volume measure $\sqrt{-g}d^4x$ and with the volume measure $\Upsilon d^4x$. To this action, we can apply the TMT procedure, which, as usual in TMT, will contain variations in all variables, including the metric tensor and the  scalars $\varphi_a$ from which $\Upsilon$ is built.

The calculation of quantum corrections and the study of possible cosmological effects caused by these  corrections to the constraint and the plateau-like potential will be considered elsewhere. But the preliminary estimates made in
    one-loop approximation show that their effect on the plateau-like shape of the potential is apparently negligible.

\acknowledgments

I am grateful to E. Guendelman, E. Nissimov and S. Pacheva for a detailed discussion of the differences between densities 
 $\Phi$ and $\Upsilon$ of the alternative volume measures (see footnote 5).

\end{document}